 \documentclass{aa}
 \usepackage{graphicx}

\begin{document}

   \title{Continuous stellar mass-loss in N-body models of galaxies}


   \author{Bruno Jungwiert \inst{1}, 
    Fran\c coise Combes \inst{2}
    and Jan Palou\v s \inst{1}}

   \offprints{B. Jungwiert \inst{1} }

   \institute{
\inst{1} Astronomical Institute, Academy of Sciences of the Czech Republic,
Bo\v{c}n\'{\i} II 1401, CZ-141 31 Prague 4\\
\email{bruno@ig.cas.cz} \\
\inst {2} DEMIRM, Observatoire de Paris, 61 Avenue de l'Observatoire,
F-75 014 Paris
}

   \date{Received 22 May 2001; accepted 4 July 2001}

\def\Mo{\rm \ M_\odot}
\def\My{\rm \ M_\odot / yr}

\abstract{
We present an N-body computer code -- aimed at studies of galactic dynamics --
with a CPU-efficient algorithm for a continuous (i.e.
time-dependent) stellar mass-loss.
First, we summarize available data on stellar mass-loss and
derive the long-term (20 Gyr) dependence of mass-loss rate of a
coeval stellar population. We then
implement it, through a simple parametric form,
into a particle-mesh code with stellar and gaseous particles.
We perform several tests of the algorithm reliability and show
an illustrative application: a 2D simulation 
of a disk galaxy, starting as purely stellar but evolving as 
two-component due to gradual
mass-loss from initial stars and due to star formation.
In a subsequent paper we will use the code to study what
changes are induced in galactic disks by the continuous gas recycling
compared to the instantaneous recycling approximation,
especially the changes in star formation rate and radial inflow of matter.
\keywords{stars: mass-loss -- 
          galaxies: evolution --
          galaxies: kinematics and dynamics --
          galaxies: spiral --
          methods: N-body simulations 
}
}

  \authorrunning{B. Jungwiert et al.}
  \titlerunning{Continuous stellar mass-loss
   in N-body models of galaxies}

   \maketitle
%

\section{Introduction}

During the last decade, many N-body simulations (e.g. Friedli
\& Benz 1993; Junqueira \& Combes 1996)
as well as theoretical works on two-fluid gravitational instabilities
(Jog 1992, 1996) have underlined the importance of the gas mass
fraction for the dynamics and evolution of galaxies.

Gas crucially conditions star formation
and evolution of large-scale instabilities (e.g. bars, spiral arms).
In turn, star formation and large-scale flows induced by these
instabilities influence mass and chemical profiles of galaxies
(e.g. Martinet \& Friedli 1997; Portinari \& Chiosi 2000).

The gas content in a given area is determined by the competing
processes of star formation and stellar mass-loss and by spatial
flows of gas.
A coherent picture of galactic evolution must therefore
consistently couple stellar and gaseous dynamics with
star formation and stellar evolution.

Star formation is, in large-scale models of galaxies, typically
implemented through simplistic parametrizations reflecting mainly
the energy supply from young stars and
the gas mass fraction locked in newly born stars.
On the other hand, stellar mass-loss has received much less attention
from galactic N-body modellers since it was for a long time considered
as a secondary issue. But nowadays, both observations and stellar
evolutionary models (see Sect. 2 for references) indicate that
the gas mass fraction restituted by stars is huge and may reach,
when integrated over the stellar mass spectrum,
some 45\% over the Hubble time. Stars thus represent not only a
place for permanent gas blocking but also an
important temporary reservoir of gas that will be gradually
reinjected into the interstellar medium.

Recent computer models of galactic dynamics usually
approximate the stellar mass-loss as instantaneous, i.e. happening at the
moment of stellar birth. Nonetheless, stars lose matter
during all their lives. Stellar lifetimes span a very large range and
in the case of low-mass stars they compare with or overpass the Hubble time.
The use of the instantaneous recycling approximation, which was
proposed by Tinsley (1980) for high mass-stars,
is therefore not satisfactory for the whole stellar mass spectrum.


From the above emerges an obvious motivation for building
a computer code able to follow the galactic structure and dynamics
together with a non-instantaneous gas recycling.
One can expect that the connection of such a recycling
with the dynamics will affect large-scale gravitational instabilities,
spatial flows of matter, star formation and gas consumption rates, etc.,
and thus the long-term evolution of galaxies.

The development and presentation of such a code
is the central aim of this paper.
Our model is innovative especially in introducing
a continuous (i.e. time-dependent) gas recycling scheme, grafted on
an underlying
N-body code using standard techniques for computing gravitational field
(particle-mesh method) and gas dynamics (sticky-particles).

The text is organized as follows. In Sect. 2, we describe data and steps
in deriving the time-dependence of mass-loss of a computer
stellar particle. This curve, simply parametrized,
is the key input information for our N-body code,
however it can be useful also for other models (not necessarily N-body) of
galaxies or star clusters. Section 3 presents our computer code, the emphasis
being put on the implementation of the mass-loss.
An instructive simulation of a barred galaxy is shown in Sect. 4.
Section 5 summarizes the outcome and future prospects.

\section{Stellar mass-loss and long-term galactic evolution}

\subsection{Mass-loss rate of a coeval stellar population}

Stars of all masses lose matter via stellar winds,
high-mass stars, in addition, in supernova explosions.
Despite the fact that stellar winds accompany all stages
of stellar evolution (see de Jager et al. 1988, for observations of
wind rates across the Hertzsprung-Russell diagram),
they vary in intensity and duration:
gas is not released uniformly over a stellar lifetime
but typically in a few relatively short episodes.

The time dependence of mass-loss rate of a coeval stellar population
(hereafter CSP),
$\dot M (t)$, could in principle be obtained if we knew
the stellar initial mass function (IMF), $\psi(m_{\rm i})$,
and the time dependence of the mass-loss rate of
individual stars, $\dot m$:
{\small
$$\dot M (t) = \int_{m_{\rm min}}^{m_{\rm max}} \dot m (t,m_{\rm i})\,
               \psi(m_{\rm i}) \, {\rm d}m_{\rm i}      \eqno(1)$$
}
Since neither the former nor the latter are known with much precision,
we shall limit ourselves to construct a model curve based on several
simplifying assumptions adequate to the purpose of the work which is
to simulate the effects of stellar mass-loss on the {\it long-term}
evolution of galaxies; we do not pretend to model changes occurring
on a time-scale of less than $\sim 10$ Myr that is short compared
to the dynamical time at most galactocentric distances.

\begin{figure} [htpb]
\resizebox{\hsize}{!}
{\includegraphics[height=4.3cm, width=4.3cm,  angle=-90]{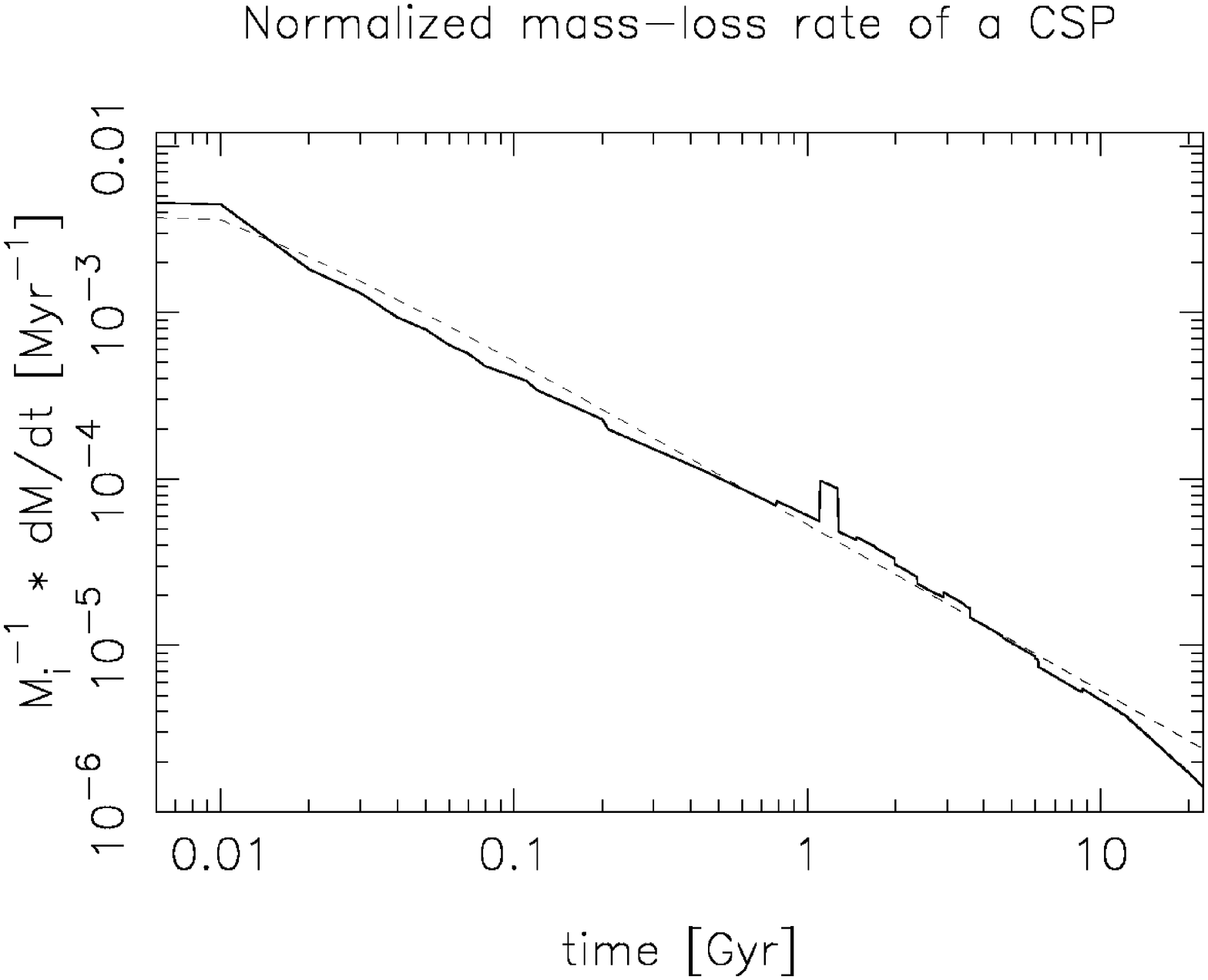}
 \includegraphics[height=4.3cm, width=4.3cm,  angle=-90]{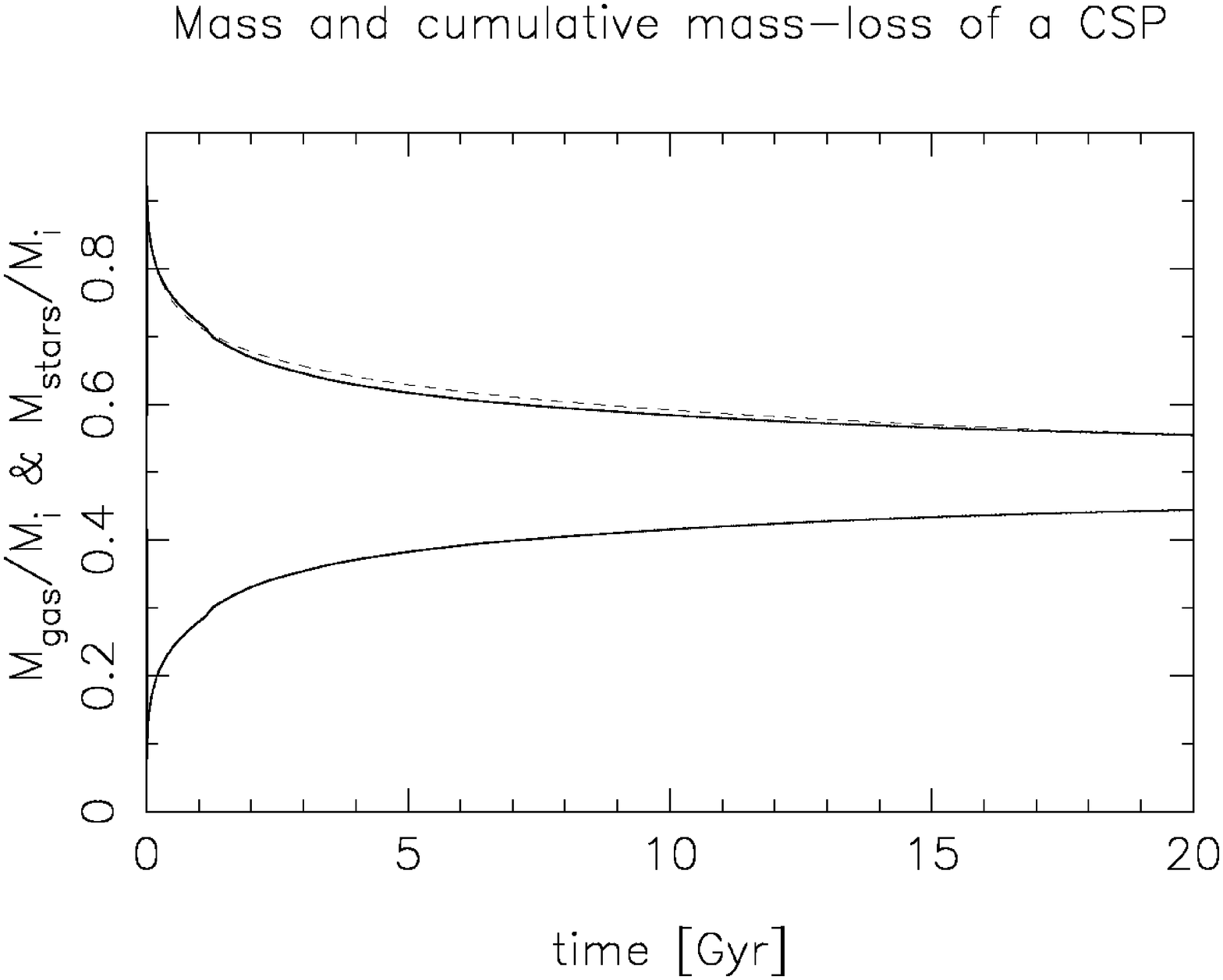}}

\caption[]{\small
{\bf Left}: Normalized mass-loss rate, $\dot M (t) / M_{\rm i} $,
of a coeval stellar population (CSP) with $Z=0.02$ and the Scalo's (1998) IMF.
The full curve is derived from the Padua stellar models, the dashed
curve is the fit corresponding to Eq. (2) with parameters from Table 1
(for $\tau=20$ Gyr).
Note that both scales are logarithmic.
{\bf Right}: Normalized stellar mass, $M_{\rm stars} / M_{\rm i}$ 
(upper full curve),
and the cumulative mass-loss (i.e. the released gas mass),
$M_{\rm gas} / M_{\rm i}$ (lower full curve), for the same CSP. The dashed 
curve
corresponds to the normalized stellar mass computed for the mass-loss rate
fit from the left pannel.}
\end{figure}

Mass-loss for stars of different masses end evolutionary stages is
reviewed in more detail in Jungwiert (1998) where references can be found
for both observations and models. Here, only a brief summary of the
most important mass-loss mechanisms is given:
a) in the case of low-mass stars (LMS, hereafter defined as
stars with initial masses $m_{\rm i}<2\, \Mo$),
winds on the red giant branch (RGB) and asymptotic giant branch (AGB)
are the most efficient;
b) for intermediate-mass stars (IMS, $2\,{\Mo} < m_{\rm i} < 8\, \Mo$), the
AGB winds dominate;
c) in the case of high-mass stars (HMS, $m_{\rm i}> 8\,\Mo$),
supernova explosions, Wolf-Rayet winds and
main-sequence winds,
contribute to the overall mass-loss in proportions strongly depending on
$m_{\rm i}$: mass-loss is dominated
by supernovae for $m_{\rm i} < 25\, \Mo$, while for more massive stars the
winds are more efficient than supernovae.

With regard to the above,
we shall assume, for simplicity, that: a) for LMS
all the mass-loss takes place in two distinct delta function events
corresponding to the tips of RGB and AGB (since the
respective winds strongly peak near these tips);
b) for IMS and HMS, we treat the mass-loss as
one delta function at the end of their lives: this reflects the AGB wind peak
for intermediate-mass stars and SN explosion + winds for high-mass stars
(as the most massive HMS,
for which the winds dominate over supernovae, have very short lifetimes
-- less than 5 Myr for $m_{\rm i} > 40\, \Mo$ -- the delta function
representation is adequate).

Necessary ingredients to construct an approximate mass-loss curve
of a CSP thus reduce to the knowledge of IMF,
stellar initial-final mass relation and
stellar lifetime-mass relation
(in the case of LMS, two lifetimes are necessary, for the ends of the RGB
and AGB phases). As for the IMF, we use the power-law with three slopes
(0.2, 1.7 and 1.3 for initial mass ranges of 0.1-1 $\Mo$, 1-10 $\Mo$
and 10-100 $\Mo$, respectively)
inferred by Scalo (1998) as the ``average'' of recently published
IMFs.
For the initial-final mass relation and lifetime-mass relation, we rely
on the Padua stellar evolutionary models (Bressan et al. 1993;
Marigo et al. 1996) computed for the
initial metallicity $Z=0.02$.

The mass-loss rate function, $\dot M_{\rm n}(t)$, constructed in this way
(the subscript ``n'' denotes the normalization to the initial mass,
$M_{\rm i}$, of a CSP) is shown in Fig. 1 (left).
For the purpose of its efficient implementation into the
N-body code, we carry out a least-squares fit.
The power-law fit, $\dot M_{\rm n} (t) \propto t^{-y}$ suggests
a very simple form since it gives $y = 1.04$, very close to unity. Rather
than using 1.04, we fix $y=1$, and perform a new least-squares fit
with a hyperbola having a shift $T_0$ from the moment $t_{\rm birth}$,
at which a CSP was born, to avoid
the unphysical singularity of the $t^{-1}$ function at $t_{\rm birth}$:
{\small
$${\dot M_{\rm n} (t)} \equiv
{\dot M(t) \over {M_{\rm i}}} = {c_0 \over {t-t_{\rm birth} + T_0}}.\eqno(2) $$
}
We restrict the least-squares fits to combinations ($c_0$, $T_0$) giving
the correct total gas return (cumulative mass-loss) $R_{\tau}$ over time
$\tau$ (see Sect. 2.2. and Table 2), corresponding to
the time interval between the birth of a given CSP and
the envisaged end of an N-body simulation.
Results of the fit for $\tau =$ 5, 10, 15 and 20 Gyr are summarized
in Table 1.

%
\begin{table}
\caption{Fitted mass-loss rate parameters}
\begin{center}\small
\begin{tabular}{ccc}
$\tau$ [Gyr]  &  $c_0$    & $T_0$ [Myr] \\
              & \\
\hbox{\ \ 5}       & $5.55 \cdot 10^{-2}$  & $5.04$ \\
10      & $5.47 \cdot 10^{-2}$  & $4.97$ \\
15      & $5.41 \cdot 10^{-2}$  & $4.92$ \\
20      & $5.35 \cdot 10^{-2}$  & $4.86$
\end{tabular} \end{center} \end{table}

The fit for the simulation length of 20 Gyr is shown in
Fig. 1 (left) by a dashed line. It is close at all times and over four orders
of magnitude on the $\dot M_{\rm n} (t)$ axis to the original dependence.
It is also
obvious, from Table 1, that the fits for different $\tau$
are very similar. We will take advantage of this when
implementing the mass-loss into the N-body code: 
only one fit will be used for all the particles (see Sects. 3.4. and
3.5.).

\subsection{Cumulative mass-loss of a coeval stellar population}

While the time-dependence of mass-loss rate of a CSP
is a vital input for our models,
the cumulative mass-loss, i.e. the mass fraction lost by stars until
time $t$ after the birth of the CSP,
$$ R_t = \int_{0}^{t} \dot M_{\rm n} (t') \,{\rm d}t',      \eqno(3)$$
gives a useful estimate of the importance of gas recycling on different
time-scales.

Fig. 1 (right) shows, in the lower half of the plot,
$R_t$ of a CSP computed
for the Scalo's (1998) IMF and the Padua $Z=0.02$ stellar models. The curve
in the upper half of the plot gives the fractional mass remaining
in stars, i.e. $1-R_t$. Table 2 gives $R_t$ for $t=$ 0.1, 1, 5, 10, 15
and 20 Gyr.

%
\begin{table}
\caption{Cumulative mass-loss of a CSP}
\begin{center}
{\footnotesize Cumulative mass-loss at several chosen times (in Gyr)
for the Scalo's IMF (1998) and Padua $Z=0.02$ stellar models.}
 \label{}
\end{center}
\vskip0mm
\begin{center}\small
\begin{tabular}{llllllllllllll}

$R_{t=0.1}$&$R_{t=1}$&$R_{t=5}$&$R_{t=10}$&$R_{t=15}$&$R_{t=20}$\\
                  &     \\
  0.170     & 0.280   & 0.382   & 0.416    & 0.434    & 0.445
\end{tabular} \end{center} \end{table}

A discussion of mass-loss for other IMFs and stellar metallicities
is beyond
the scope of this paper. $R_t$ for several other IMFs widely used in
the litterature (Salpeter 1955; Miller \& Scalo 1979; Kennicutt 1983)
as well as for the Padua stellar models corresponding to
$Z$=0.0004, 0.004, 0.008, 0.02, 0.05 is tabulated in Jungwiert (1998).
Here we note only a few facts. For $Z=0.02$, the highest gas return
over the Hubble time (let's say 15 Gyr) is
obtained for the Scalo's (1998) and Kennicutt's (1983) IMFs
($R_{15\, \rm{Gyr}} = 0.43$ and 0.44, respectively),
followed by the Miller \& Scalo's (1979) IMF (0.41), while
the Salpeter's (1955) IMF gives a markedly lower value (0.30).
In all the cases, the returned mass fraction is considerable.
As for the relative contributions of LMS, IMS and HMS to the
total mass-loss, they are comparable, so that no category
can be considered as significantly dominant.
For example, the Scalo's (1998) IMF and $Z=0.02$ stellar models
lead to ratios LMS : IMS : HMS $\sim 1:1.25:1$ for the contributions
to $R_{15\, \rm{Gyr}}$. The stellar metallicity does not change the mass-loss
of the whole CSP too much despite the fact
that individual stars have their lifetimes and final
masses influenced by Z quite noticeably (see Figs. 16 and 17b
in Jungwiert 1998).

\subsection{Galactic gas return rate}

We define, for the purpose of simulations presented in Sect. 4,
{\it gas return rate} ($GRR$) as the analog of star
formation rate ($SFR$): $GRR$ (units of $\Mo\, yr^{-1}$) 
is the rate at which
stellar mass is converted into gaseous mass. In general,
$GRR$ depends on the whole history of star formation:

$$ GRR\,(t) = \int_0^t \, SFR\,(t') \ \dot M_{\rm n} \,(t-t')\, {\rm d}t',
\eqno(4)$$
where $\dot M_{\rm n}\, (t-t')$ is the (normalized)
mass-loss rate, at time $t$, of a stellar population born at $t'$.
Note that when $GRR$ refers to a specified area, $SFR$ in Eq. (4)
does not refer to the same area: $GRR$ at time $t$ is not determined
by the star formation history of this area but by the star formation
history of stellar matter that is now present in this area but has,
in general, various orbital histories.

\section{N-body code with star formation and time-dependent
stellar mass-loss}

\subsection{Gravitational interaction and equations of motion}

We use the standard particle-mesh (PM) technique 
(Hockney 1970; Hohl 1971; for a review, see: Hockney \& Eastwood 1981; 
Selwood 1987) to compute the
gravitational potential on a Cartesian grid via Fast Fourier Transform
(FFT).
Forces at grid points are then evaluated by differencing the
potential. Both the assignment of particle masses to grid points and
the finding of forces between them are carried out by the bilinear
``cloud-in-cell (CIC)'' interpolation. The equations of motion are
integrated by means of the leap-frog algorithm 
(Hockney \& Eastwood, 1981).

The gravitational softening
that reduces interparticle forces at short distances 
(see Pfenniger \& Friedli 1993) is that of a Plummer sphere. 
The choice of the softening length is done with regard
to the criteria laid down by Romeo (1994, 1997) to optimize the stability
and relaxation properties of N-body disks.

\subsection{Particle species}

There are four particle species in our code. Apart from
relatively massive stellar particles (hereafter referred to as
``standard stars''), typically representing millions
of stars, and relatively massive gaseous particles
(hereafter ``clouds'') with characteristic masses of
$10^5-10^6 \Mo$, we introduce low-mass stellar and low-mass
gaseous particles (hereafter nicknamed ``starlets'' and
``cloudlets'') that mediate the exchange of matter -- converted
from gaseous to stellar phase and vice versa by star formation
and stellar mass-loss -- between the standard stars and clouds.
The definition of mass boundaries
between cloudlets and ordinary clouds and between starlets and
ordinary stars, and thereby the mass ranges for cloudlets and starlets, 
are precised in Section 4.1. (item 13). 

The standard clouds are meant to simulate the cold gas component.
They are modelled as finite-size particles
that undergo physical collisions in which they dissipate a fraction
of kinetic energy. The relative velocity of two
colliding clouds is reduced by factors $\beta_{\rm r}$, $\beta_{\rm t}$; 
the former
multiplies the velocity component along the line joining the two
clouds (radial component), the latter perpendicular to it (tangential
component) (for pioneering use of this ``sticky-particles'' technique
in simulations of galaxies, see Schwarz 1981; among more recent
applications and discussion on values of $\beta_{\rm r}$, $\beta_{\rm t}$, see
e.g. Jungwiert \& Palou\v s 1996).

Star formation is allowed to proceed in our cold gas component
in the way described in Sect. 3.3. It results in the production
of the above mentioned starlets. These in turn join the standard stellar
component by merging with a standard star provided some is
near by (typically 200 pc; to be precised below). On the other
hand, standard stars lose continuously mass in the form of the
cloudlets -- as detailed in Sect. 3.4. -- that join the
cold gas component by merging with a standard cloud provided some
is near by. The condition of particle ``nearness'' for star-starlet
or cloud-cloudlet merging ensures that the mass exchange between
the stellar and gaseous components is local.

In usual situations
the existence of our starlets and cloudlets will be ephemeral since
they will merge
with a nearby standard particle immediately after
they are born. This is technically equivalent to a direct exchange
of a fraction of mass between a standard stellar and a standard
gaseous particle. However, there are important and frequent
astrophysical situations in which stars and cold gas are segregated
spatially. For example,
in some areas cold gas can be depleted by intense star formation
or swept away by gravity torques. There can also be regions of
cold gas without any or with only small amounts of stars,
for instance the whole galaxies in early evolutionary stages
or, in the case of present-day galaxies, the area beyond
the periphery of optical disks.

Our starlets/cloudlets enable
to follow such situations. If a starlet/cloudlet is produced
in an area (defined below) with no standard star/cloud,
it leads an independent and possibly long-term
existence until it meets the respective standard particle
(meanwhile, the starlet is subject to mass-loss as standard stars
and it produces cloudlets).
The long-living starlets/cloudlets can also meet
another starlet/cloudlet. Such events are
treated by starlet-starlet and cloudlet-cloudlet mergings.
In the course of time,
a starlet/cloudlet grown by successive mergings
of starlets/cloudlets might approach the mass
typical of standard stars/clouds. If this occurs, the starlet/cloudlet
is converted into a standard star/cloud.
Conversely, a standard cloud is changed over to a cloudlet if
its mass falls, due to the starlet production, below the value typical of
standard clouds. In contrast, we do not consider a conversion of 
standard stars into starlets since the mass of a standard star remains 
of the same order during a simulation: it cannot decrease 
below  
the quantity $1-R_{t_{\rm end}}$ that is larger than 
$\sim 0.55$ for any simulation length $t_{\rm end}$ shorter than 20 Gyr 
(see Table 2). 

In this respect, the cloudlets can be viewed as a
very crude representation of the warm/hot galactic gas: they are
products of star formation and do
not participate in the star forming process unless they ``cool''
in areas of high gas density (either by merging with a standard cloud or
with many other cloudlets).

In summary, the four particle species have the following
properties. {\it Standard stars} produce cloudlets and can absorb
starlets. {\it Starlets} can be absorbed by standard stars or merge
with other starlets; they produce cloudlets; if their mass
exceeds a given value ($M_{\rm sS}$, $\sim 10^5 - 10^6 \Mo$, see Sect. 4.1.,
item 13),
they are converted into standard stars.
{\it Standard clouds} undergo mutual,
partly inelastic collisions; they produce starlets and can absorb
cloudlets; if their mass falls below
a given value ($M_{\rm cC}$, $\sim 10^5 \Mo$, Sect. 4.1., item 13),
they are converted into cloudlets. {\it Cloudlets} can be absorbed
by standard clouds or merge with other cloudlets; if their mass
exceeds the above mentioned $M_{\rm cC}$,
they are converted into standard clouds. 

All the particle
productions, absorptions, mergings and collisions happen in the
mass and momentum conserving manner.
 
Perhaps the most questionable interaction in our code 
is the merging of a starlet with a standard star or another starlet.
The real stellar fluid cannot dissipate energy; nevertheless,
if new stars are born with a low velocity dispersion 
(as expected from the kinematics of parent gas clouds), 
the velocity dispersion of the real stellar fluid as a whole can 
decrease. We merge the particles for computational reasons, so as to keep
their number low (see below); we are unable to follow every new born stellar 
particle independently. In this context, the merging of standard stars 
with starlets can be viewed as very crudely mimicking the rejuvenation of the
stellar fluid by newly born stars and the related decrese of its 
velocity dispersion.
The merging of one starlet with another can seem more problematic but
this process is relatively rare in simulations we present. However,
we admit that the merging of stellar particles could in general bias
the evolution of the stellar orbital structure, especially if the merged
particles had very different velocities. This problem certainly merits
further investigation.

As indicated above, the interparticle interactions
(collisions, mergings) require the ability to identify
nearby particles. The same is true for the star formation process
described in the subsequent subsection.
To find nearby particles in a CPU-efficient way,
we divide a galaxy into a 2D rectangular grid (hereafter
``interaction grid''; for its specifications see Sect. 4.1., 
items 11 and 6); four lists (standard stars list, starlets
list, standard clouds list and cloudlets list) are associated
with it: within each of them, the particles are continuously sorted
in such a way that their indices are stored in the order of the
increasing index of the grid boxes. This permits a fast
identification of all the particles in a given box at every time
step. Particles located in a given box are considered as
``nearby'' in the above used sense.
It is true that two particles from distinct neighbouring boxes may be closer
to each other than particles within a same box. However, our concern
is basically to have at most one starlet and one cloudlet per box and to
achieve this in a computationally simple manner. The fact of ignoring
interactions behind the box boundary is dynamically unimportant since
it happens on distances below the dynamical resolution of the code.

The star-starlet, starlet-starlet, cloud-cloudlet and cloudlet-cloudlet
mergings permit to keep a low number of particles.
If starlets/cloudlets were not merged, they would soon proliferate
in an inacceptable number since at each time step
at which the star formation and mass-loss procedures are applied,
roughly $N_{\rm boxes}$ ($\sim 10^4 - 10^5$) starlets/cloudlets are created.
The merging rule usually leads to mergings of nearly all
starlets/cloudlets with standard stars/clouds within
the time step in which these starlets/cloudlets were created
(see Sect. 4.3.) except areas or evolutionary stages where or
when there is not enough stars or cold gas.

\subsection{Star formation scheme}

We impose a generalization of the Schmidt law for star formation
(Schmidt 1959; for a recent review, see Kennicutt 1998), i.e. the
surface density of star formation rate, $\mu_{SFR}$, is proportional
to a power of the gas surface density:
$\mu_{SFR} = A \, \mu_{\rm gas}^{\alpha}$.

 Kennicutt (1998) gives $\alpha=1.4$ as the
``best-fit'' for the observationally determined Schmidt law
relating the mean surface density of $SFR$
to the mean surface density of cold
(molecular + atomic) gas in both galactic disks and circumnuclear regions
of disk galaxies.
In our model, we keep the freedom of choosing $\alpha$ as an input parameter
(typically 1 or 2). The Schmidt law is applied with a time step
$\Delta t_{\rm SF}$ (longer than the time step used to
integrate the equations of motion; Sect. 4.1., item 12) separately
in each of the boxes of the interaction grid.

Inside a given box, star formation proceeds as follows:
1) The standard clouds list is used to compute the cold gas surface density.
2) If this density is non-zero (i.e. there is at least one
standard cloud),
a starlet is created with the mass computed according to the Schmidt law.
3) The starlet mass is subtracted from the standard clouds
in amounts proportional to their masses.
4) The position and velocity of the starlet are found in such a way that
the total momentum of the standard clouds and of the starlet is conserved.
5) Standard stars, if any, are found using their list.
If there are some, one of them is randomly chosen to merge with
the starlet. If there is no standard star, the starlets list is searched 
through for present starlets, if any, that
immigrated from neighbouring boxes.
If the search is positive, all the starlets in the box merge
into one. Otherwise the new starlet will temporarily lead an independent
existence as described in the previous subsection.

\subsection{Stellar mass-loss scheme}

In Sect. 2 we have evaluated the normalized
mass-loss rate, $\dot M_{\rm n} (t)$, of a coeval stellar population.
Here our goal is to implement the mass-loss into the N-body code.

As described in two previous subsections, new stellar mass
forms, in our code, from gas as starlets that can merge with
nearby standard stars or other starlets.
Our typical stellar particle (either a standard one or a starlet)
therefore does not represent a CSP but has a complex star formation history
since it consists of mass of starlets born at various times in addition to
the  mass it was assigned in the beginning of the simulation.

The exact treatment of the mass-loss -- for the code purposes discretized
in time with a step $\Delta t_{\rm ML}$ (Sect. 4.1., item 12) --
would then require to keep in memory, for each stellar particle,
its star formation history, i.e. initial masses of starlets that were
incorporated into it and the times at which these starlets were born.
The mass lost by stellar particle $J$ at time $t_l$ due to incorporated
starlets
would read:
{\small
$$\Delta M_J (t_l) = \int_{t_l-\Delta t_{\rm ML}}^{t_l} \,
  \sum_{k<l} m_{*J} (t_k) \cdot \dot M_{\rm n} (t' - t_k)\, {\rm d}t' =$$
$$=\sum_{k<l} m_{*J} (t_k) \cdot \Delta M_{\rm n} (t_l - t_k),\eqno(5)$$
}
where $m_{*J} (t_k)$ is the mass with which a starlet, incorporated into $J$
at $t<t_l$, was created at time $t_k$ (the incorporation could have happened
later than at $t_k$),
and $\Delta M_{\rm n} (t_l - t_k)$ is the fraction of mass lost
over $t_l-\Delta t_{\rm ML}$
by the stellar mass born at $t_k$. Computing this sum for every
stellar particle ($N_{\rm stars+starlets}\sim 10^5$)
in each step ($N_{\rm steps} \sim 1000$ if mass-loss is applied every
$\Delta t_{\rm ML} \sim 10$ Myr
over the Hubble time) and keeping in memory the
arrays of $m_*$ and $t_k$ would be computationally very expensive.

We therefore introduce a much simpler manner to evaluate the
mass-loss of stellar particles getting benefit from the fact that
$\dot M_{\rm n} (t)$ was shown to be reasonably well fitted by a simple
analytical function with only two parameters (Eq. 2).
The trick consist in using the
same functional form for all stellar particles disregarding their
star formation histories, while individually adjusting the two parameters
whenever a new starlet is incorporated.

It is thus necessary to define how to change the relevant parameters
when two stellar particles of different ages and masses merge,
and to check whether or not such a change plausibly
represents the sum of their mass-loss rates.
Formulated mathematically, we want to find, for a particle $J$ formed
at time $t_J$ by merging of particles $R$ and $S$, born at $t_R$ and
$t_S$, parameters $C_J$ and $T_J$ such that the mass-loss rate of $J$ (hereafter
called the {\it combined} mass-loss rate), expressed as
{\small
$$\dot M_J (t) = {C_J \over {t-t_J + T_J}}    ,\eqno(6) $$
}
would have similar time dependence and total gas return as if particles
$R$ and $S$ were followed independently. Eq. (5)
is then replaced by the corresponding integral of eq. (6):
{\small
$$\Delta M_J (t_l) =
- C_J\, \ln(1-{\Delta t_{\rm ML}\over {t_l - t_J + T_J}})
\eqno(7)  $$
}
Since there are two free parameters, $C_J$ and $T_{J}$,
we can impose two conditions on the combined mass-loss.
After trying several possibilities 
(for a more detailed analysis, see Jungwiert (1998), pages 52-58) 
we require Eq. (7) to give
the correct cumulative mass-loss at two time moments: at the end of
the simulation, $t_{\rm end}$, and at an intermediate time, $t_{\rm im}$,
between $t_J$ and $t_{\rm end}$; the choice of
$t_{\rm im} = t_J + 0.2 \cdot (t_{\rm end} - t_J)$ proved to give good
results. These two conditions can be expressed as
{\small $$ \int_{t_J}^{t_{\rm x}} \dot M_J (t) \,{\rm d}t =
   \int_{t_J}^{t_{\rm x}} \dot M_R (t) \,{\rm d}t +
   \int_{t_J}^{t_{\rm x}} \dot M_S (t) \,{\rm d}t,                 \eqno(8)$$}
where $t_{\rm x}$ denotes either $t_{\rm end}$ or $t_{\rm im}$.

A priori, it was not ensured that our combined mass-loss approximation
well represented the real sum of particles' mass-losses. However, the
tests presented in Sect. 3.6. verify the adequacy of our approach.

Knowing how to combine the mass-loss rates of individual particles when
they merge, it remains to be said what values of $C_J$ and $T_J$
we assign to particles that do not form by merging, i.e.
to starlets at the moment of their birth and to standard stars, if any,
present in the beginning of the simulation.
Such can be viewed as representing coeval stellar populations.
Comparing eqs. (2) and
(6), we see that $T_J$ is simply $T_0$ and $C_J$ is $c_0$ multiplied by
the mass of a particle at the moment of its birth,
$t_{\rm birth}=t_J$.
Values of $c_0$ and $T_0$ are chosen from Table 1 (as we have
mentioned in Sect. 2.1., the fits presented in Table 1 are very
close to each other; we will use that for $\tau = 10$ Gyr).

For the standard stars present in the beginning
of the simulation, there is an additional issue consisting in
defining their $t_{\rm birth}$; we postpone it
to a special subsection below.

The last point is to explain what happens with masses 
$\Delta M_J$ lost by
individual stellar particles at every time step $\Delta t_{\rm ML}$. We
use a strategy similar to that for star formation (Sect. 3.3.). 
The procedure
runs through the interaction grid box by box:
1) All the standard stars and starlets are found using the
respective lists and their $\Delta M_J$ (computed
according to Eq. (7)) are summed up. 2) If the sum is non-zero
(i.e. there is at least one stellar particle in the box), the
mass is attributed to a new cloudlet. 
3) It is positioned  
to the center of mass of contributions $\Delta M_J$ 
coming from individual stellar particles and its
velocity is computed in such a way that the total momentum
of stellar particles in the box and of the cloudlet itself is conserved.
4) The standard clouds, if any, are found using the clouds list.
If there are some, one of them is randomly chosen to merge with
the cloudlet. If there is no standard cloud,
the cloudlets list is searched through for cloudlets, if any,
that immigrated from neighbouring boxes.
If the search is positive, all the cloudlets in the box merge
into one. Otherwise the new cloudlet will temporarily lead an independent
existence.

\subsection{Mass-loss of pre-existing stars}

The purpose of this subsection is two-fold: first, to drive attention
to the fact that when a simulation of a galaxy starts from a state
in which a stellar component is already present, as is the case of numerous
published works, it is natural and important to take into account
its mass-loss; second, to define $t_{\rm birth}$ and $C_J$ for
standard stellar particles representing this component
(hereafter ``pre-existing stars'' and ``pre-existing stellar component'').

To know at what rate the pre-existing stellar component should lose mass
during the simulation, an assumption must be
made about the star formation history of the model galaxy. Here we
restrict our attention to two extreme cases, a constant star formation
rate over some epoch
prior to the start of the simulation (defined by $t=0$) and a starburst
at some moment
prior to the start of the simulation.
To be more specific, we suppose a simulation of length $t_{\rm end}=10$ Gyr
and calculate what gas mass, $\Delta M_{\rm gas}$, is released {\it
during the simulation} (i.e. between $t=0$ and $t_{\rm end}$)
by the pre-existing stellar component. Table 3 gives this
mass (relative to the mass of the pre-existing stellar population at
$t=0$) for two constant $SFR$ histories (lasting 5 and 10 years) and
two starbursts (occuring 5 and 10 years prior to the start of the
simulation):
%
\begin{table}
\caption{Gas released by pre-existing stars}
\begin{center}\small
\begin{tabular}{lc}
pre-simulation SF: &
$\Delta M_{\rm gas} (t_{\rm end} =10\, {\rm Gyr}) \over{M_{\rm stars} (t=0)}$  \\
                       &     \\
const. $SFR$ (-5 to 0 Gyr)  &  0.153  \\
const. $SFR$ (-10 to 0 Gyr) &  0.115  \\
starburst (-5 Gyr)       &  0.084 \\
starburst (-10 Gyr)      &  0.050
\end{tabular} \end{center} \end{table}
%

The values of $\Delta M_{\rm gas}$ range between 5 and 15\%
of the pre-existing stellar mass. This can represent a huge amount
of gaseous matter in absolute terms. For example, starting a simulation
with a stellar component of $4\cdot 10^{10}\, \Mo$ (as will be the case
of our illustrative simulation in Sect. 4) implies that the gas mass
released from these stars between $t=0$ and $t_{\rm end} = 10$ Gyr would range
between $2\cdot 10^9\, \Mo$ and $6\cdot 10^9\, \Mo$ which is
typical of gas mass in disks of observed present-day spiral galaxies.
The mass-loss of pre-existing stars is thus quite important and should
not be neglected in simulations.

Technically, to be able to use Eq. (7), we need to assign to each
of the pre-existing star (having mass $M_0$ at $t=0$)
a birth time, $t_{\rm birth}=t_{J}$, and a birth mass, $M_{\rm birth}$
(the latter will give us $C_J$ after multiplying by $c_0$).
In the starburst case, all such particles have $t_{\rm birth}$
simply equal to the time moment of the starburst; in the case of
a constant $SFR$ history, we draw $t_{\rm birth}$ for each pre-existing
particle randomly within the period of constant star formation epoch.
Once we assign $t_{\rm birth}$, $M_{\rm birth}$ is found by integrating Eq. (2)
backwards in time (from $t=0$ to $t_{\rm birth}<0$).

Pre-existing stars are thus supposed to have already lost mass
$M_{\rm birth} - M_0$ prior to the start of the simulation,
however their mass-loss will continue also during the simulation.

\subsection{Tests}

Several tests of our mass-loss algorithm are described in Jungwiert (1998).
Especially, it has been tested that the combined mass-loss (Sect. 3.4.)
of a stellar particle built by merging (``composite particle'' for
the purpose of this subsection) of two or more stellar particles
(``merged particles'')
well approximates the sum of individual mass-losses for
many mass and age ratios of merged particles and for many subsequent
mergings. Five examples are shown in Figs. 2-4. In each row,
the left pannel shows the mass-loss rate of a composite particle
while the right pannel shows its mass and cumulative mass-loss
(i.e. the gas mass it has released). Full lines represent quantities
computed as if all the merged particles
were followed independently (i.e. according to Eq. 5) while the dashed
lines correspond to the combined mass-loss approximation (eqs. 6 - 8)
used in our code. All the quantities are normalized to the
sum of initial masses of merged particles.

\begin{figure} [htpb]
\resizebox{\hsize}{!}
{\includegraphics[height=4.3cm, width=4.3cm,  angle=-90]{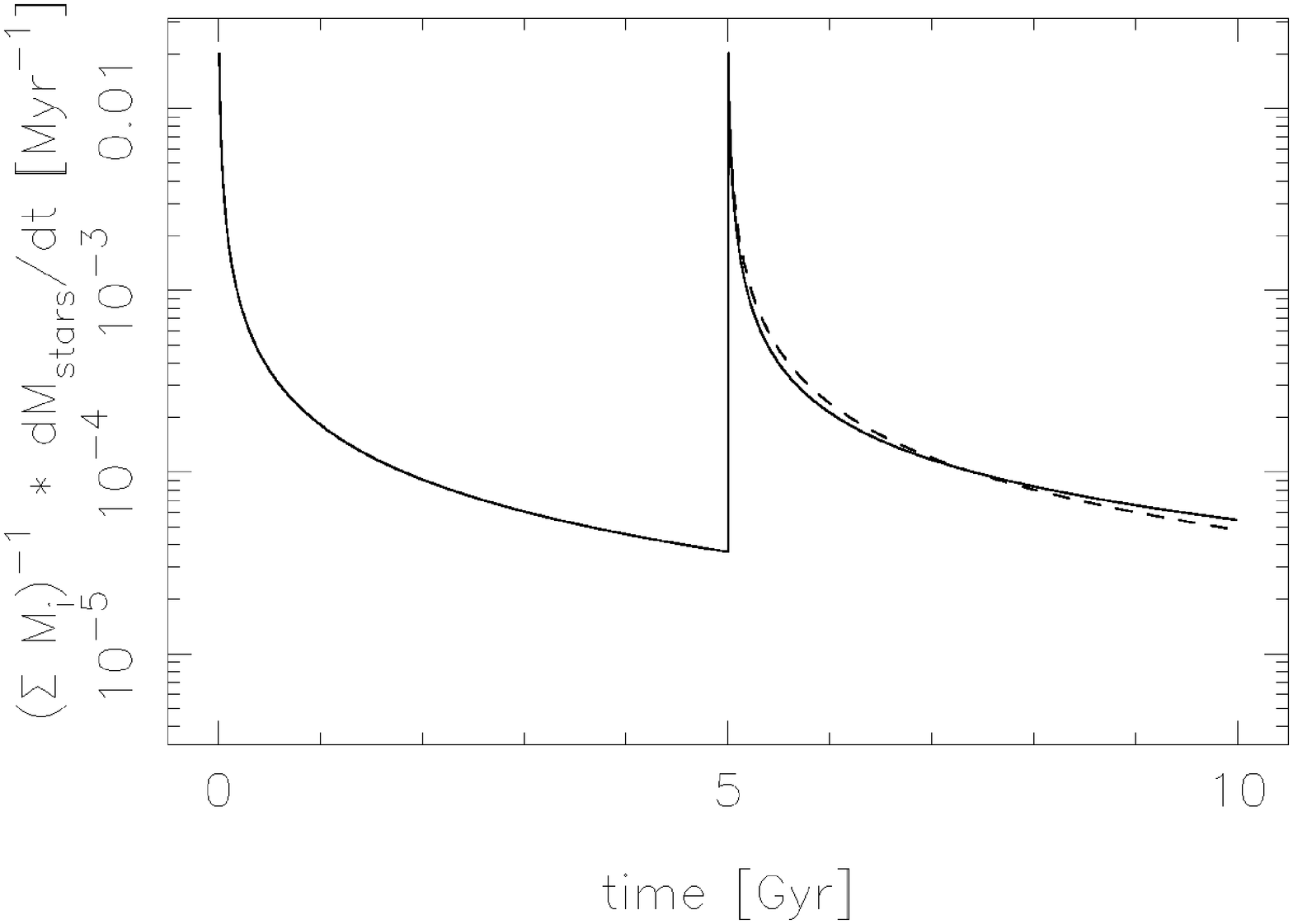}
 \includegraphics[height=4.3cm, width=4.3cm,  angle=-90]{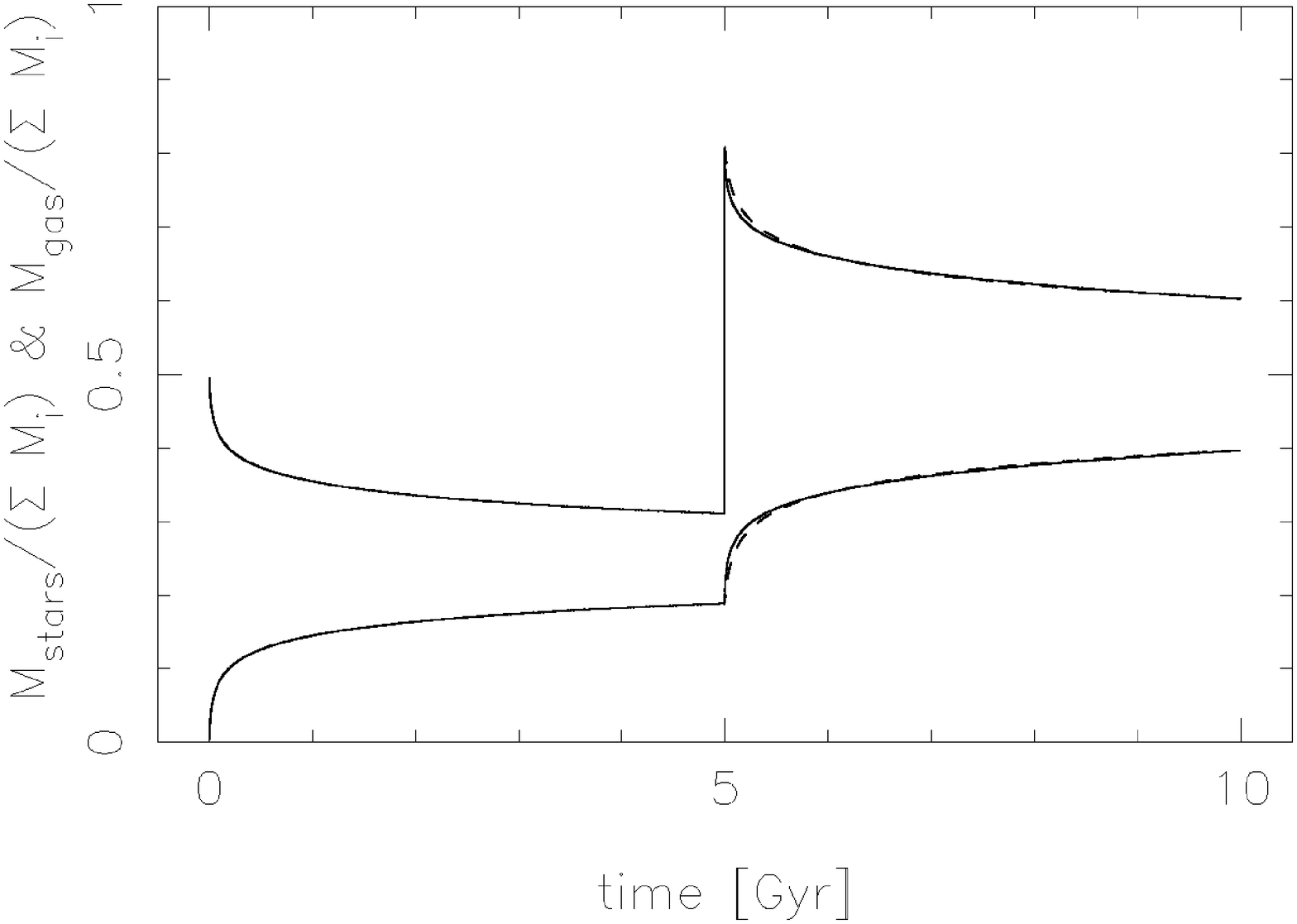}}

\caption[]{\small
Merging of two stellar particles born at $t=0$ and
5 Gyr. {\bf Left}: mass-loss rate of the composite particle;
{\bf right}: mass ($M_{\rm stars}$, upper curve) and cumulative mass-loss
($M_{\rm gas}$, lower curve) of the composite particle.
Full lines in Figs. 2-4 correspond to exact sums of mass-losses
of individual particles, dashed lines represent our combined mass-loss
approximation.}
\end{figure}

 \begin{figure} [htpb]
{\includegraphics[height=4.3cm, width=4.3cm,  angle=-90]{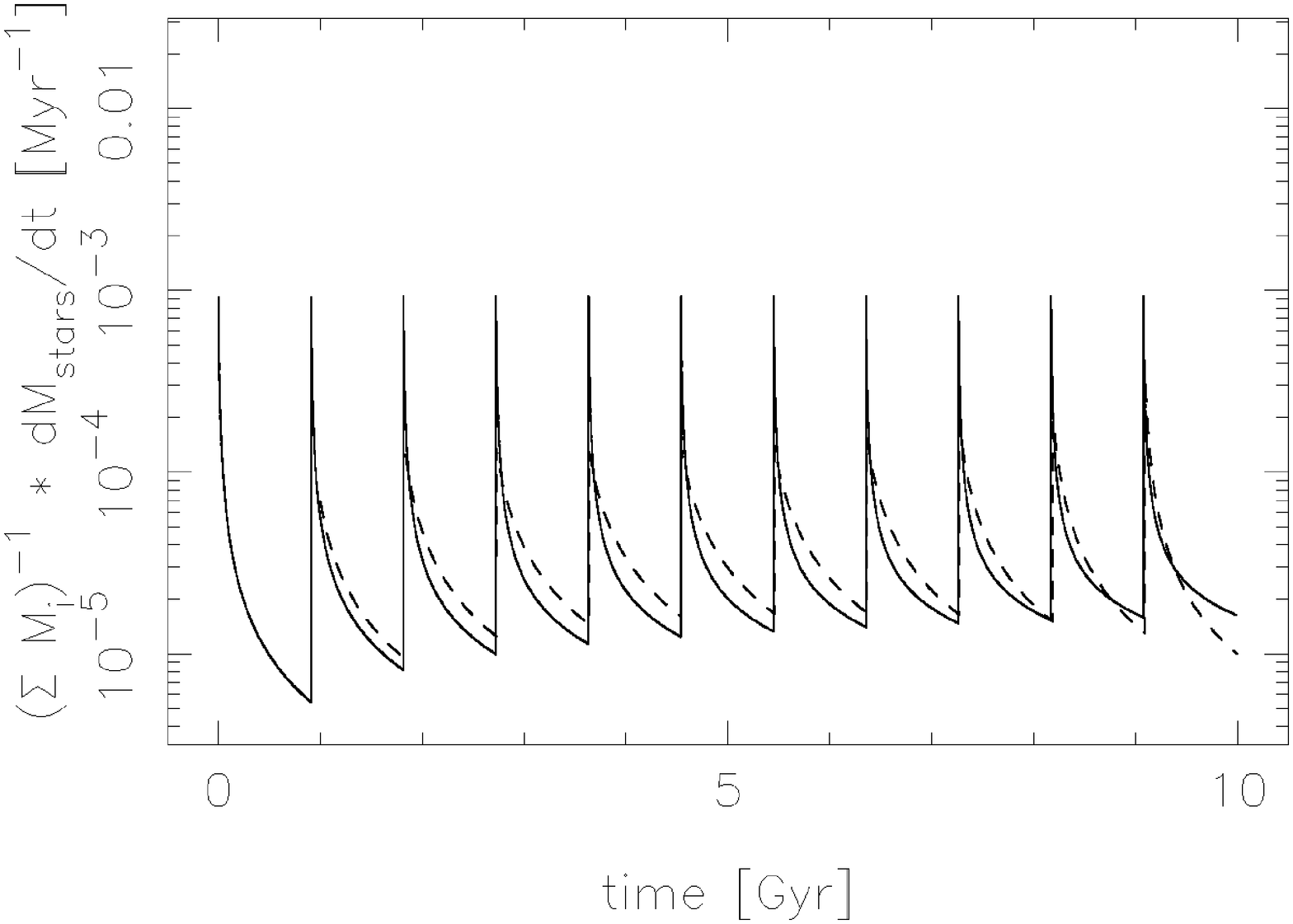}
 \includegraphics[height=4.3cm, width=4.3cm,  angle=-90]{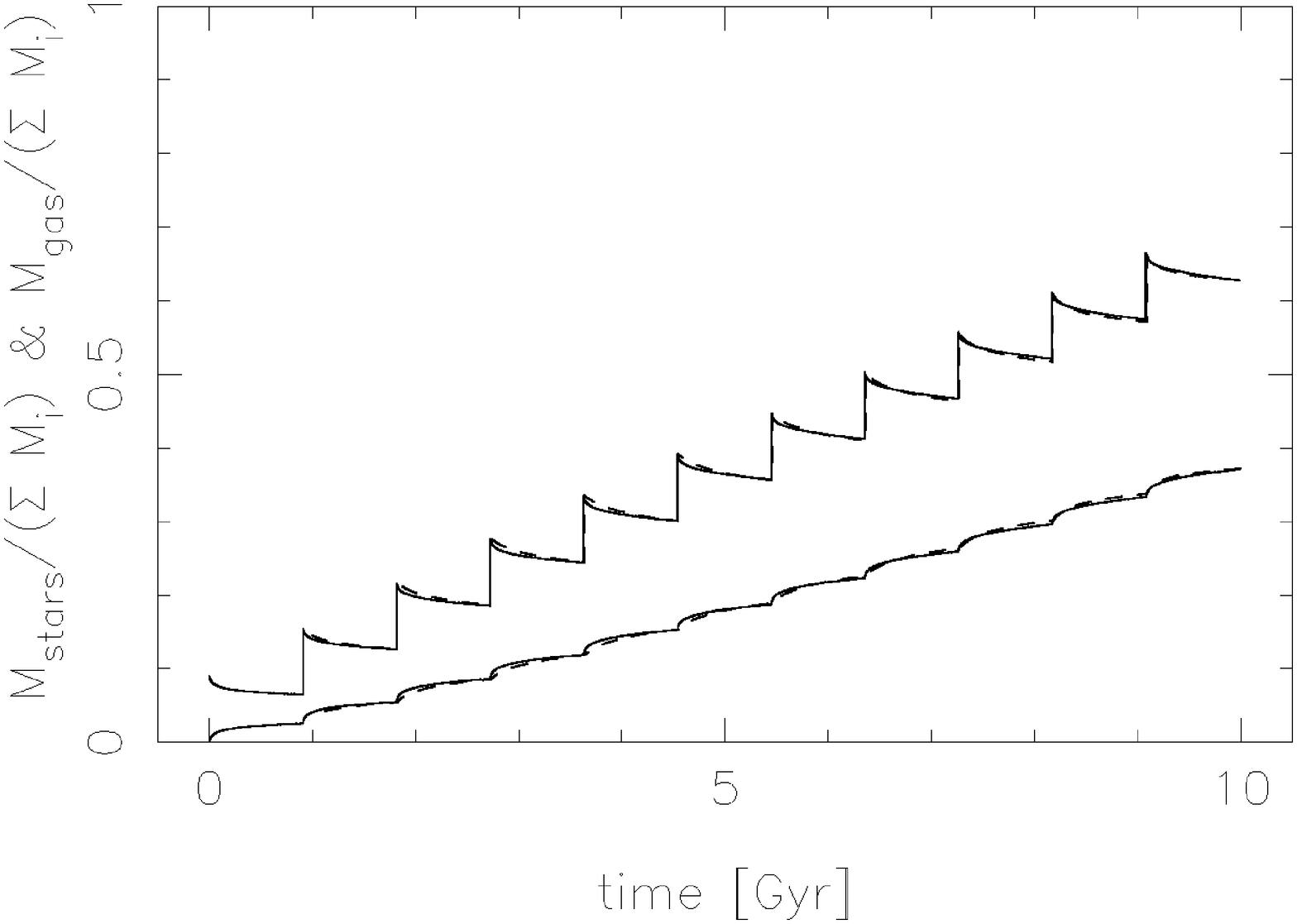}}
\vskip3mm
{\includegraphics[height=4.3cm, width=4.3cm,  angle=-90]{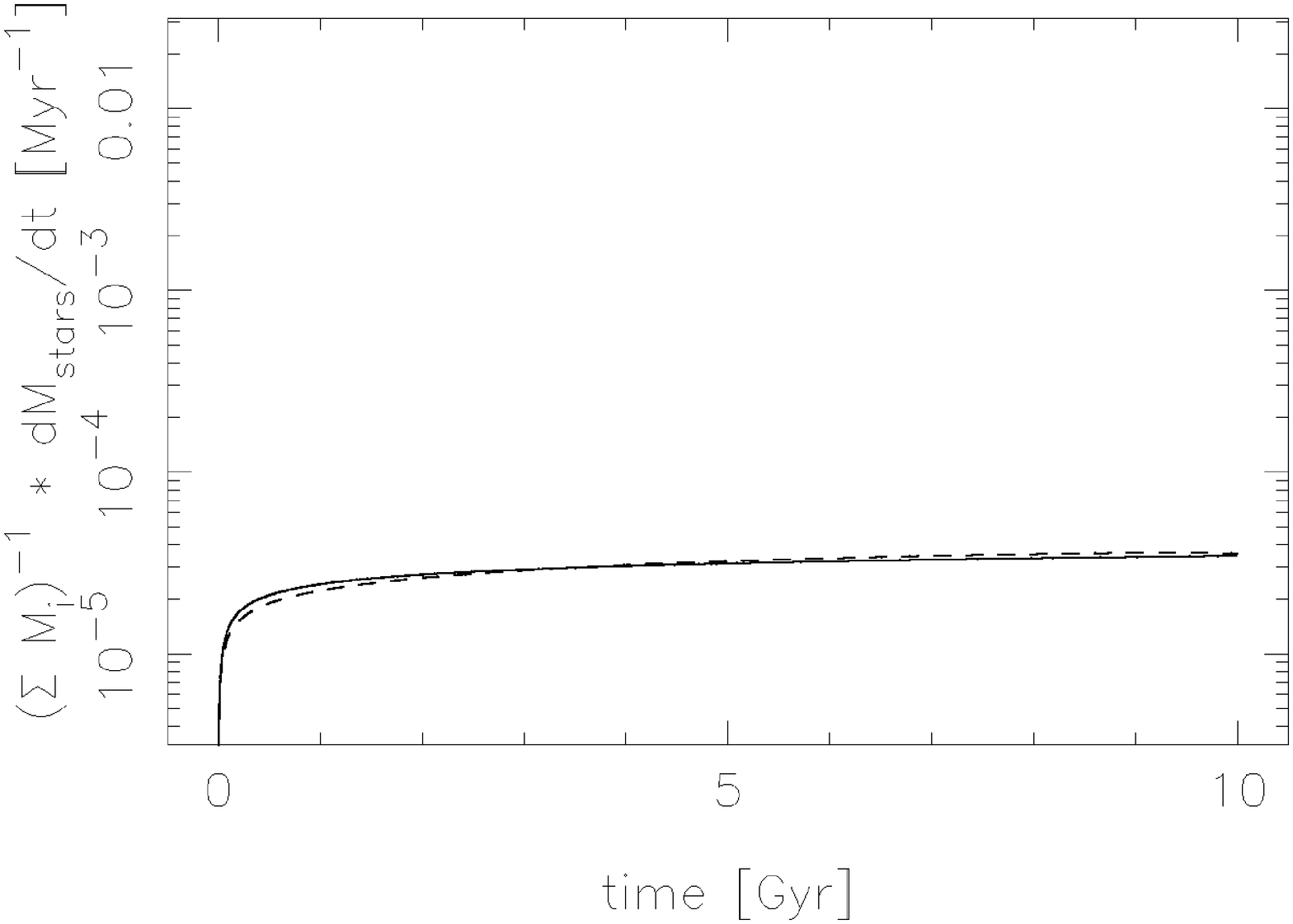}
 \includegraphics[height=4.3cm, width=4.3cm,  angle=-90]{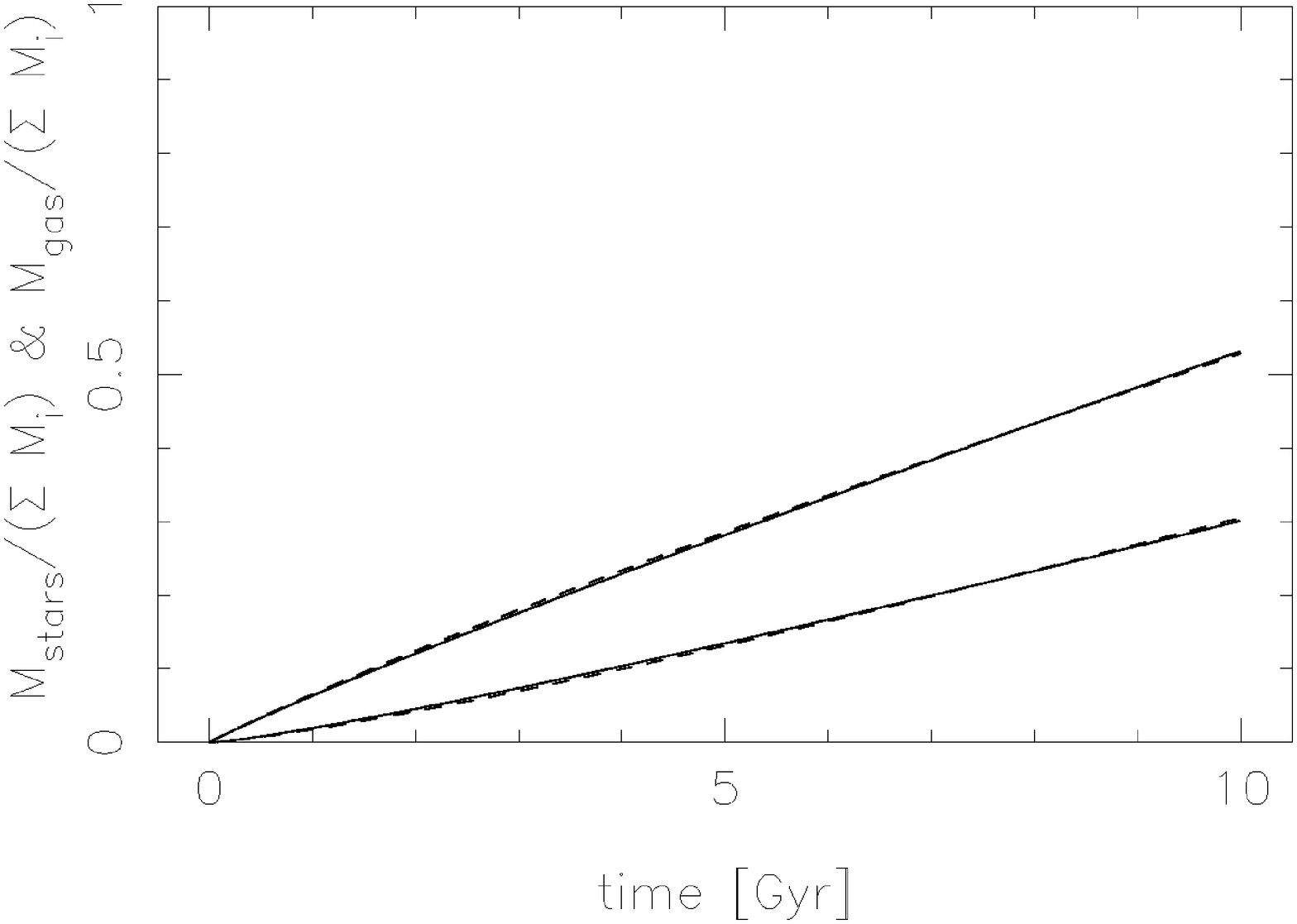}}

\caption[]{\small Like Fig. 2 but for multiple mergings of particles
having the same initial mass and born equidistantly in time. 
The particles merge at their birth.
{\bf Top}: $N_{\rm merge} =10$;
{\bf bottom}: $N_{\rm merge} =10000$.}
\end{figure}

\begin{figure} [htpb]
{\includegraphics[height=4.3cm, width=4.3cm,  angle=-90]{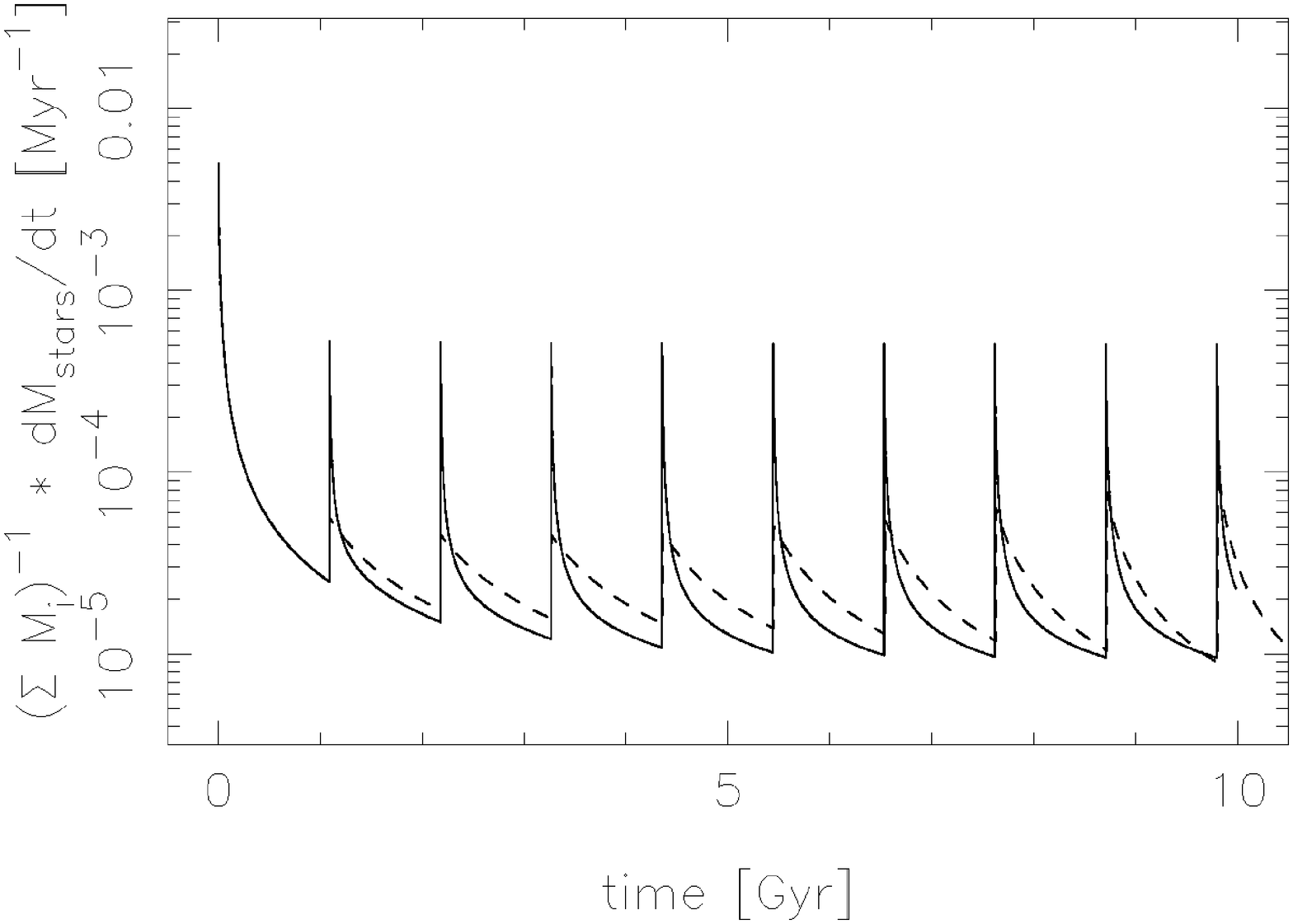}
 \includegraphics[height=4.3cm, width=4.3cm,  angle=-90]{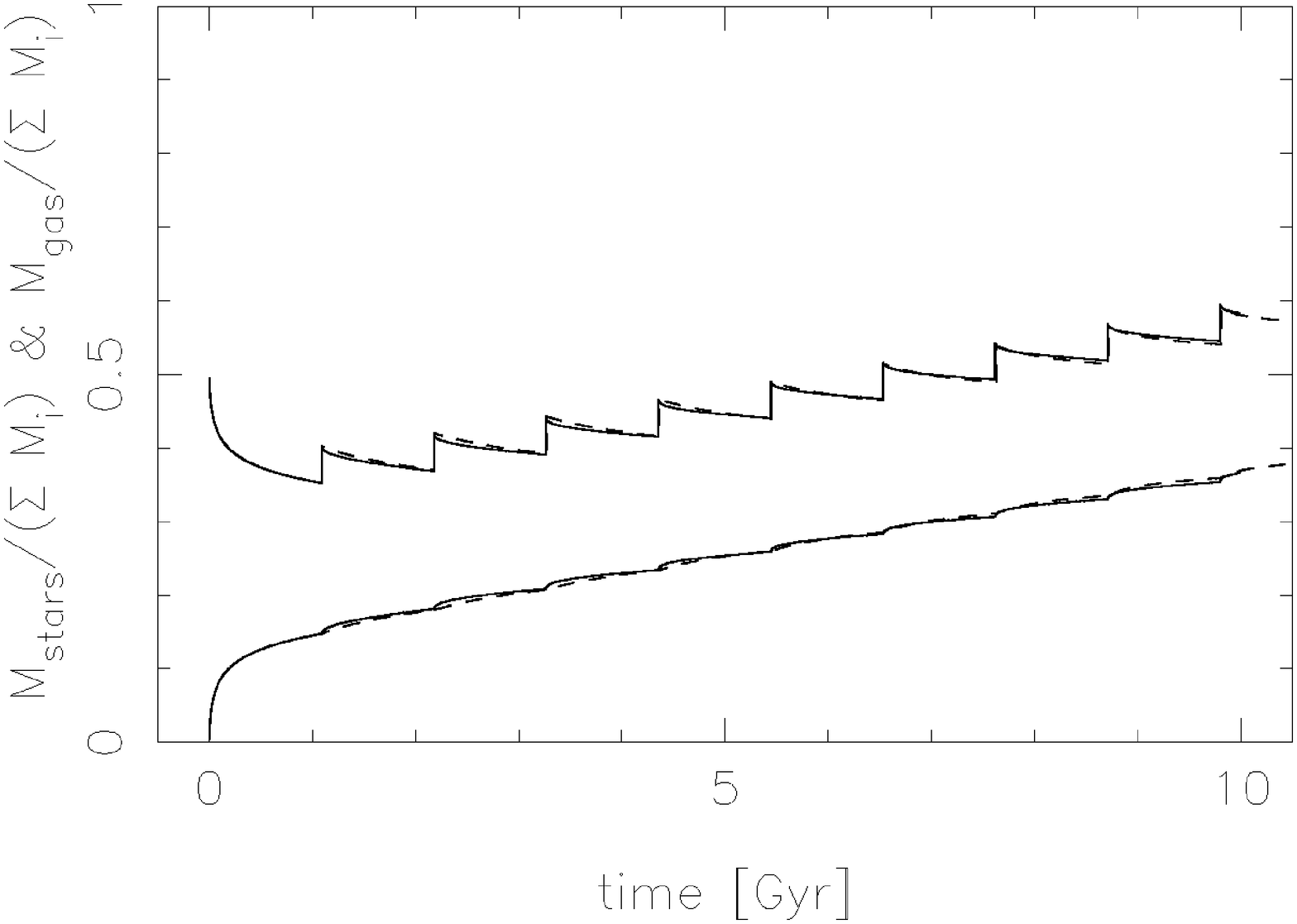}}
\vskip3mm
{\includegraphics[height=4.3cm, width=4.3cm,  angle=-90]{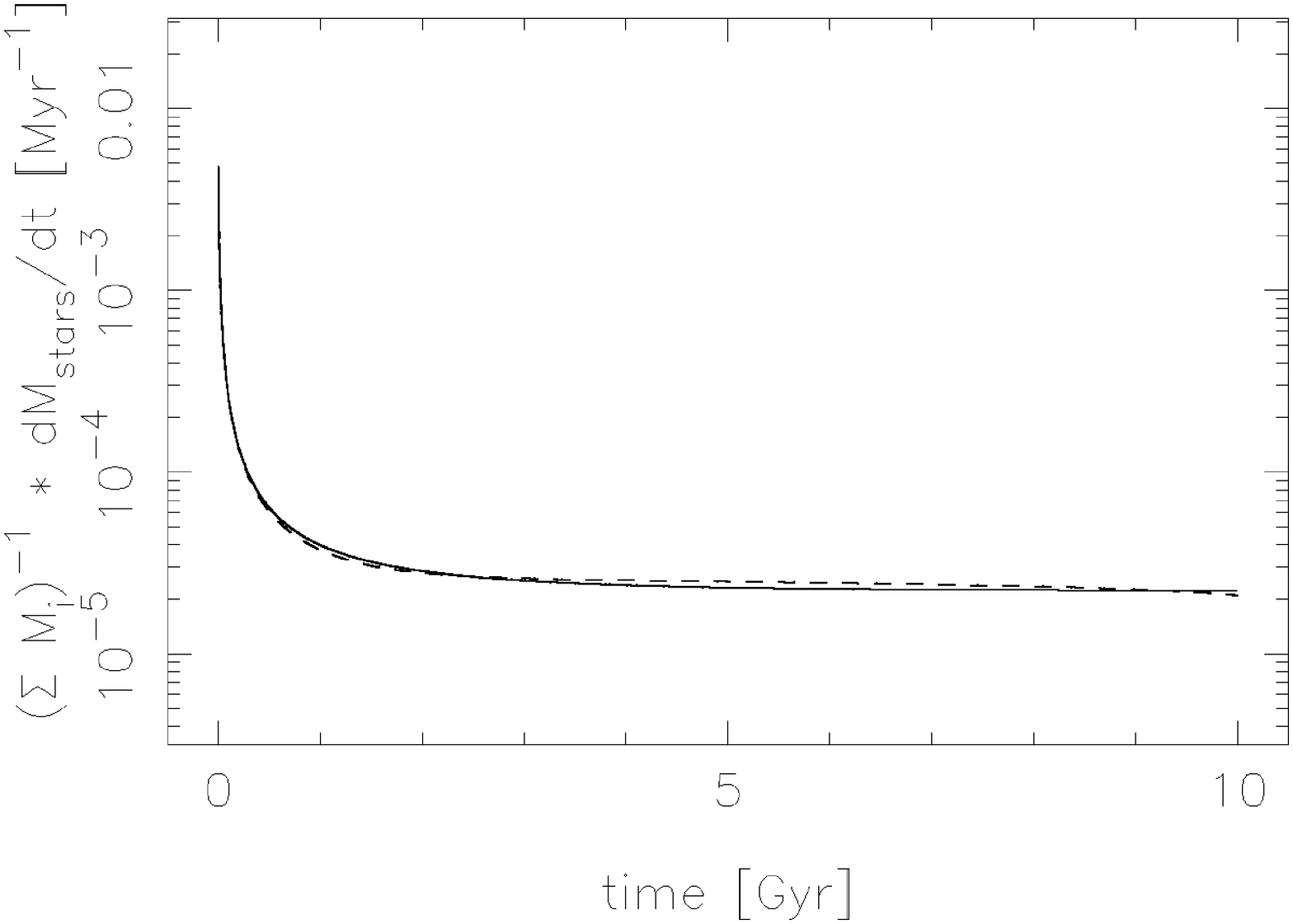}
 \includegraphics[height=4.3cm, width=4.3cm,  angle=-90]{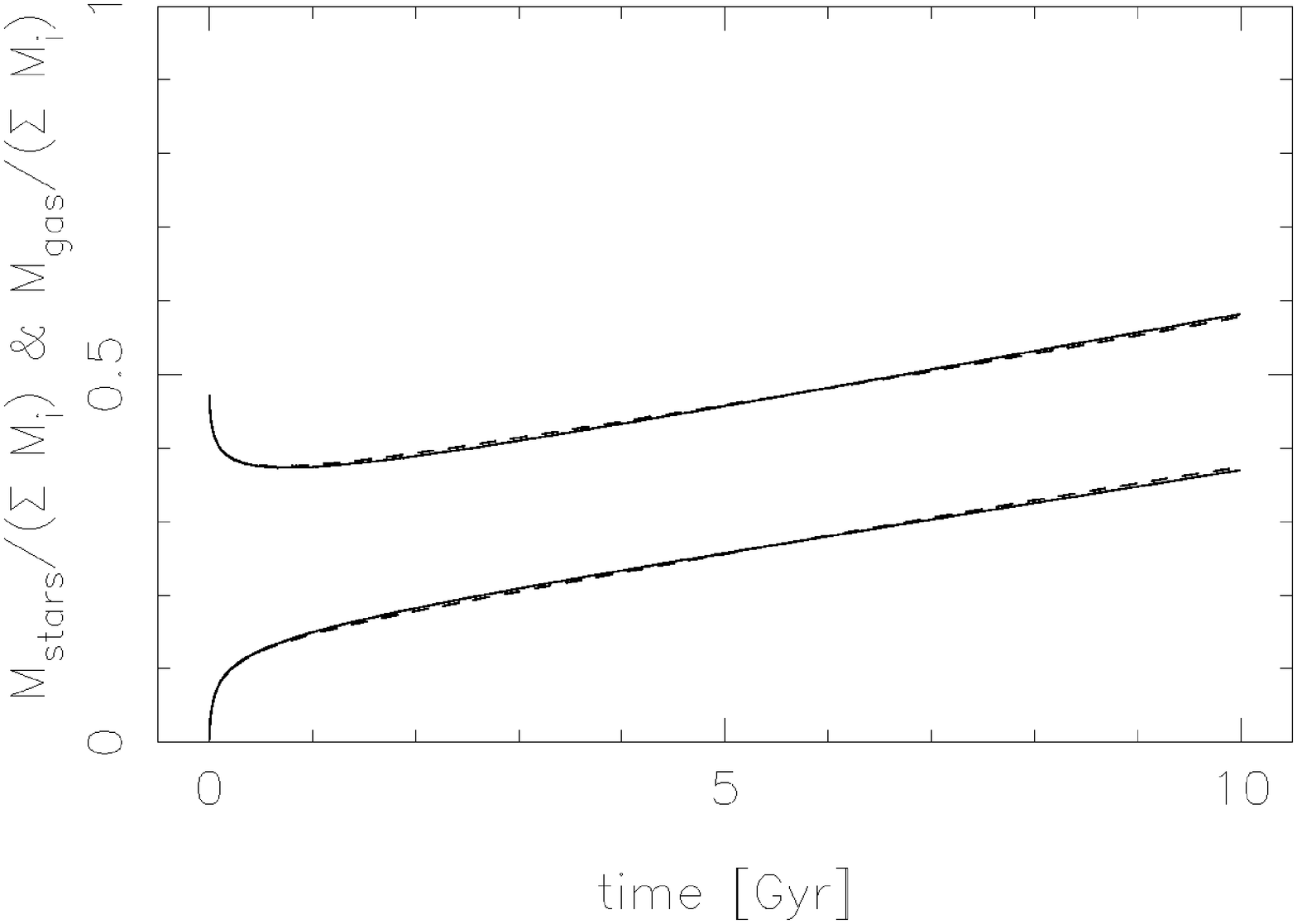}}

\caption[]{\small Like Fig. 3 (i.e multiple mergings of particles
born equidistantly in time) but the initial mass of the first particle,
$m_{\rm i,1}$, is higher than initial masses, $m_{\rm i,others}$,
of other particles it merges with.
{\bf Top}: $N_{\rm merge} =10$, $m_{\rm i,1}=10\, m_{\rm i,others}$;
{\bf bottom}: $N_{\rm merge} =10000$, $m_{\rm i,1}=10000\, m_{\rm i,others}$.}

\end{figure}

Fig. 2 corresponds
to a merging of two stellar particles of the same initial mass.
The first one is born at $t=0$, the second at $t=5$ Gyr when the
merging takes place. Fig. 3 shows the case of multiple mergings, equidistant
in time, with all the merged particles having the same initial mass.
The number of particles, $N_{\rm merge}$, that have merged into the composite
particle is 10 in the top row and 10 000 in the bottom row. Fig. 4
also corresponds to $N_{\rm merge} = 10$ (top row) and 10 000 (bottom row)
but now the first particle is ten times,
respectively ten thousand times more massive than the others it merges with.
In all the examples presented in Figs. 2-4 the combined mass-loss
approximation well represents the exact mass-loss. A few additional
examples can be found in Jungwiert (1998, pages 52-58).

We have also carried out
several runs of the N-body model in which the motion of particles was
inhibited while the star formation and mass-loss schemes, producing
cloudlets and starlets and combining them with standard
clouds and stars, were operational. We have verified that these runs
yield temporal evolutions for $SFR$ and $GRR$ close to numerical
solutions of models not using any particles.

\section {An example: a disk galaxy}

As an illustrative application of our code, we present
4 simple 2-D models of a disk galaxy having no gas in the
beginning of simulations. The first two models
(denoted A and B) are rather artificial
but useful reference cases: in model A stellar
mass-loss is supressed, the disk remains 100\% stellar (classical
purely stellar N-body model); in model B mass-loss from
initial stellar particles
produces gas but star formation is supressed. The standard models
(denoted C1 and C2)
are those in which both star formation and stellar mass-loss
operate: model C1 uses the linear Schmidt law ($\alpha=1$),
model C2 the quadratic one ($\alpha=2$);
stellar mass-loss is due both to initial
and newly formed stellar particles. Table 4 summarizes basic
attributes of the models.
The simulation length is 10 Gyr for all of them.

%
\begin{table}
\caption{Basic attributes of models}
\begin{center}\small
\begin{tabular}{cccc}
      &  & star formation & mass-loss \\
Model &  &              & \\
A     &  &   no         & no         \\
B     &  &   no         & yes        \\
C1    &  &   yes, $\alpha=1$ & yes \\
C2    &  &   yes, $\alpha=2$ & yes
\end{tabular} \end{center} \end{table}

\subsection{Input data}

For the sake of completness and reproducibility, we enumerate,
before describing the simulations, the code input parameters
and specify their numerical values. The majority of them will be
common to a larger set of models that will be published in a
separate paper. Items 1-7 refer to parameters characteristic
of collisionless N-body models, items 8-13 are specific to our
gas dynamics and star formation/mass-loss schemes.

1) Initial stellar surface density (2D Toomre-Kuzmin disk:
$M_{\rm disk} = 4 \cdot 10^{10} \Mo$,
scale-length of 4 kpc, truncation radius of 16 kpc);

2) Initial number of stellar particles, $N_{\rm stars} = 50000$; 
all of them have initially the same mass 
$M_0 = M_{\rm disk} / N_{\rm stars} = 8\cdot 10^5 \Mo$.

3) Initial velocity distribution. We fix the Toomre (1964) 
stability parameter 
$Q_{\rm stars} = 1.5$ (constant in space), and then obtain
velocities of particles by finding the time-independent axisymmetric
solution of the Jeans equations;  the ratio of azimuthal-to-radial 
velocity dispersions is computed using 
the epicyclic approximation (see e.g. Binney \& Tremaine 1987);   

\begin{figure*} [htpb]
\centering
{\includegraphics[width=18cm]{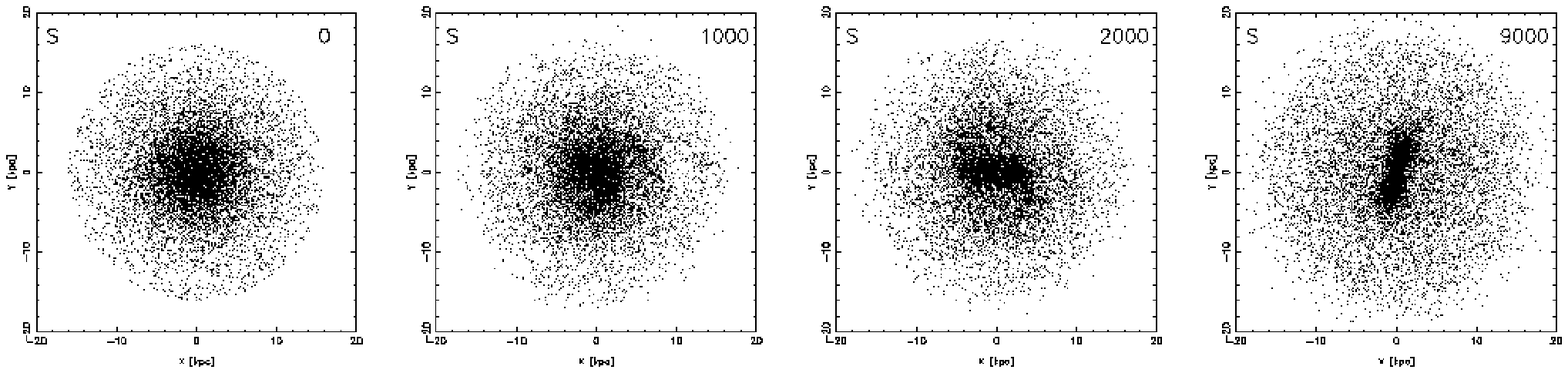}}
\vskip-2mm

\caption[] {\small Disk evolution in the purely stellar model A.
In Figs. 5-7: numbers in the top right corner give time in Myr; 
only 20\% of standard stellar particles, i.e. 10000, are shown 
for clearness.}
\vskip3mm
{\includegraphics[width=18cm]{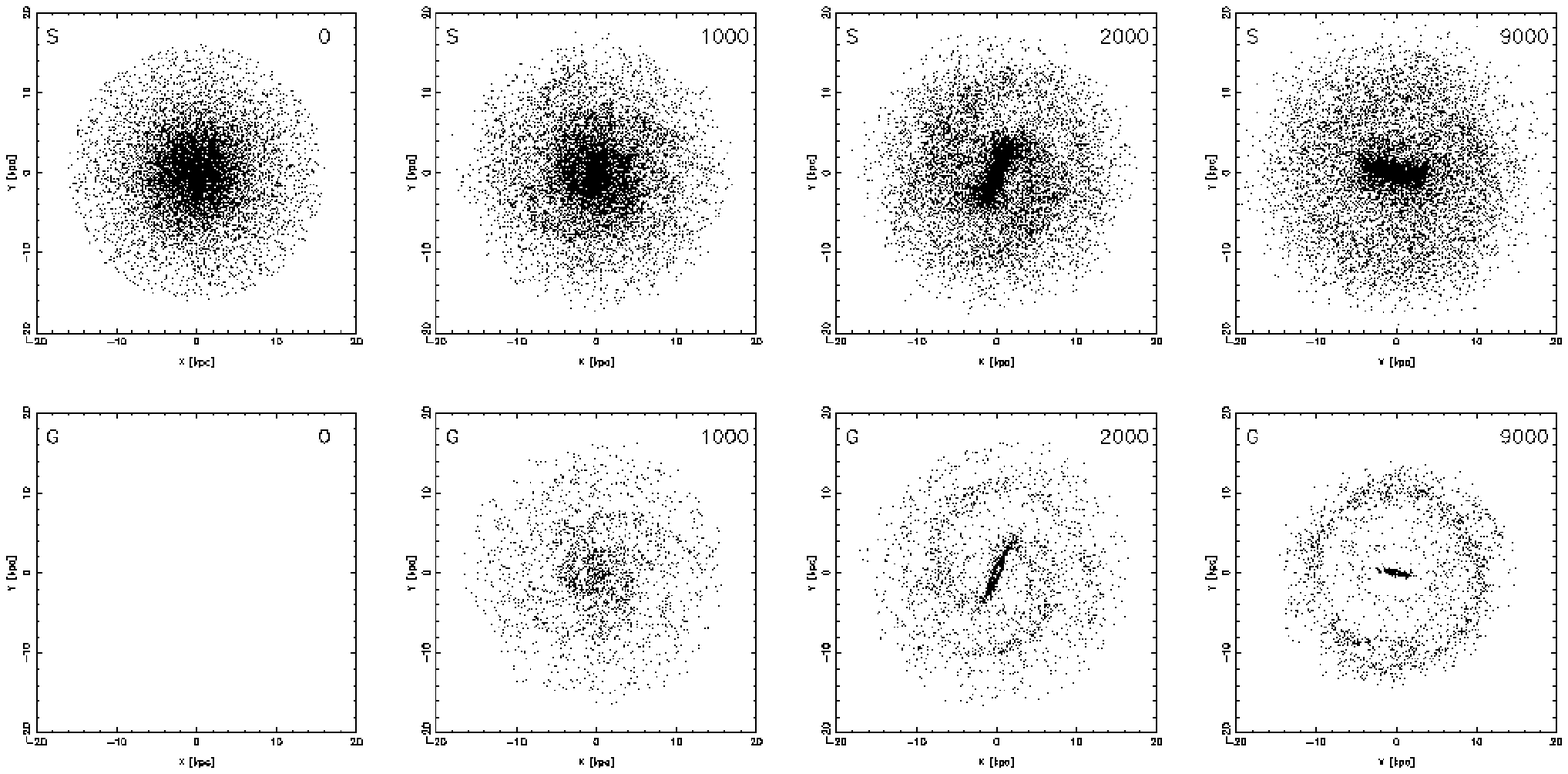}}
\vskip5mm
{\includegraphics[width=18cm]{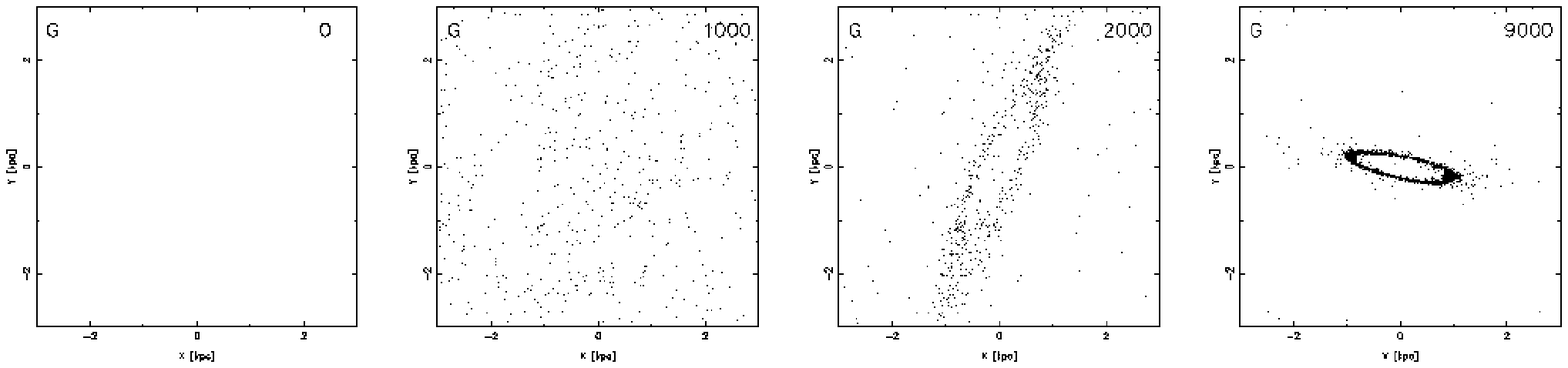}}
\vskip-2mm
\caption[] {\small Disk evolution in the standard model C1.
{\bf Top}: standard stars; 
{\bf middle}: standard clouds; {\bf bottom}: 
standard clouds -- zoom to the central region.}

\end{figure*}

4) Bulge and halo masses ($M_{\rm bulge} = M_{\rm disk}$,
$M_{\rm halo} = 5\cdot 10^{11} \Mo$) and scale-lengths (Plummer spheres of
1 and 25 kpc, respectively). The bulge is implemented essentially
to partly stabilize the inner disk, the halo to keep the rotation curve
flat in the outer disk. Both components (introduced by
analytical formulae) are treated as unevolving;

5) Gravitational softening length (Sect. 3.1.), $s \approx 0.2$ kpc. 
It corresponds
to the Romeo's (1994, 1997) ``safe'' softening computed for the center
of our disk.

6) Number of divisions of the Cartesian grid used to compute
the gravitational potential (Sect. 3.1.): $ND = 256$ (in each dimension).
This number relates to the ``active'' grid
(the computation proceeds on a grid doubled in each dimension as required
by the FFT technique).
The linear size of a grid cell, $\Delta_{\rm grid}$, is set equal to
the softening length. The above choice of $ND$ and $s$ gives the
linear size of the active grid, $D_{\rm grid} = 50$ kpc
(allowing for maximum disk radius of 25 kpc);

7) Time step for the integration of equations of motion, $\Delta t$.
It is constant over the whole disk. Its value varies with time in such a
way that it is 1/25 of the minimum orbital period. Initially
$\Delta t = 0.6$ Myr. It decreases as the central mass concentration
increases;

8) Parameters for cloud collisions (Sect. 3.2.): a) $\beta_{\rm t}=1.0$. This
choice assures the conservation of angular momentum in a collision; b)
$\beta_{\rm r}=0.6$.
This choice is rather arbitrary. Varying $\beta_{\rm r}$ falls
beyond the scope of this paper. We just note that a positive sign for
$\beta_{\rm r}$ means, in our notation, that the direction of the cloud-to-cloud
momentum in a collision is not reversed, unlike in most
works implementing this collisional scheme. For more comments on values
of $\beta_{\rm r}$, see Jugwiert \& Palou\v s (1996);

9) Two parameters for the Schmidt star formation law (Sect. 3.3.):
we choose $\alpha = 1$ or $2$ and
compute the proportionality constant $A$ in such a way that
the $SFR$ value coincides with that of the Kennicutt's (1998) fit
($\alpha=1.4$) at cold gas surface density of $5 \Mo\, pc^{-2}$
(a value typical of present day-spiral galaxies);

10) Parameters related to the stellar mass-loss rate (Sects. 3.4.
and 3.5):
to assign $t_{\rm birth}$ to initial stellar particles,
we use a simplifying arbitrary assumption that $SFR$ was constant
over a period $T_{\rm SF} = 5$ Gyr before the beginning of the simulation.
$c_0$ and $T_0$ (that determine $C_J$ and $T_J$ of individual
standard stars and starlets) are taken from Table 1 
(for $\tau=t_{\rm end}=10$ Gyr);

11) The interaction grid (Sect. 3.2.) used for
cloud collision, star formation and stellar mass-loss schemes
has the same size and the same number of cells as the grid described
in item 6;

12) Time steps $\Delta t_{\rm SF}$ (Sect. 3.3.) and
$\Delta t_{\rm ML}$ (Sect. 3.4.) for applying star formation and mass-loss
schemes. We fix $\Delta t_{\rm SF} = \Delta t_{\rm ML} = 6$ Myr (10 times the
initial integration step $\Delta t$);
 
13) Masses $M_{\rm sS}$ and $M_{\rm cC}$ for star-starlet and cloud-cloudlet 
conversions (Sect. 3.2.). 
We fix $M_{\rm sS} = (1-R_{t=10\, {\rm Gyr}}) \cdot M_0 \sim 3.3\cdot 10^5 \Mo $, 
i.e. equal to the minimum mass to which the cumulative mass-loss $R_t$
(Table 2) can drive a stellar particle with initial mass $M_0$. As for
$M_{\rm cC}$, we set $10^5 \Mo$; the overwhelming part of cold gas
in galaxies seems to dwell in clouds with higher masses than this value.

\subsection{Results}

The disk evolution in the purely stellar model A and in the standard 
model C1 is shown in Figs. 5 and 6, respectively.
Stellar disks (Fig. 5 and upper row of Fig. 6), initially axisymmetric, 
first develop transient spiral arms and then a long-lasting bar.   
Meanwhile, in model C1, stars release gas in the form of cloudlets. These
gradually merge and in regions of high gas density form standard clouds 
(see Sect. 3.2.) that 
kinematically react to the non-axisymmetric potential and 
accumulate predominantly in a dense nuclear ring (Fig. 6, bottom row) and
in an outer ring (Fig. 6, middle row), 
connected respectively to the inner and outer Lindblad
resonances (ILR, OLR). A hint of another, weaker gaseous ring,
surrounding the stellar bar
and related to the 4/1 resonance, can also be seen.

\begin{figure} [htpb]
\includegraphics[width=9cm]{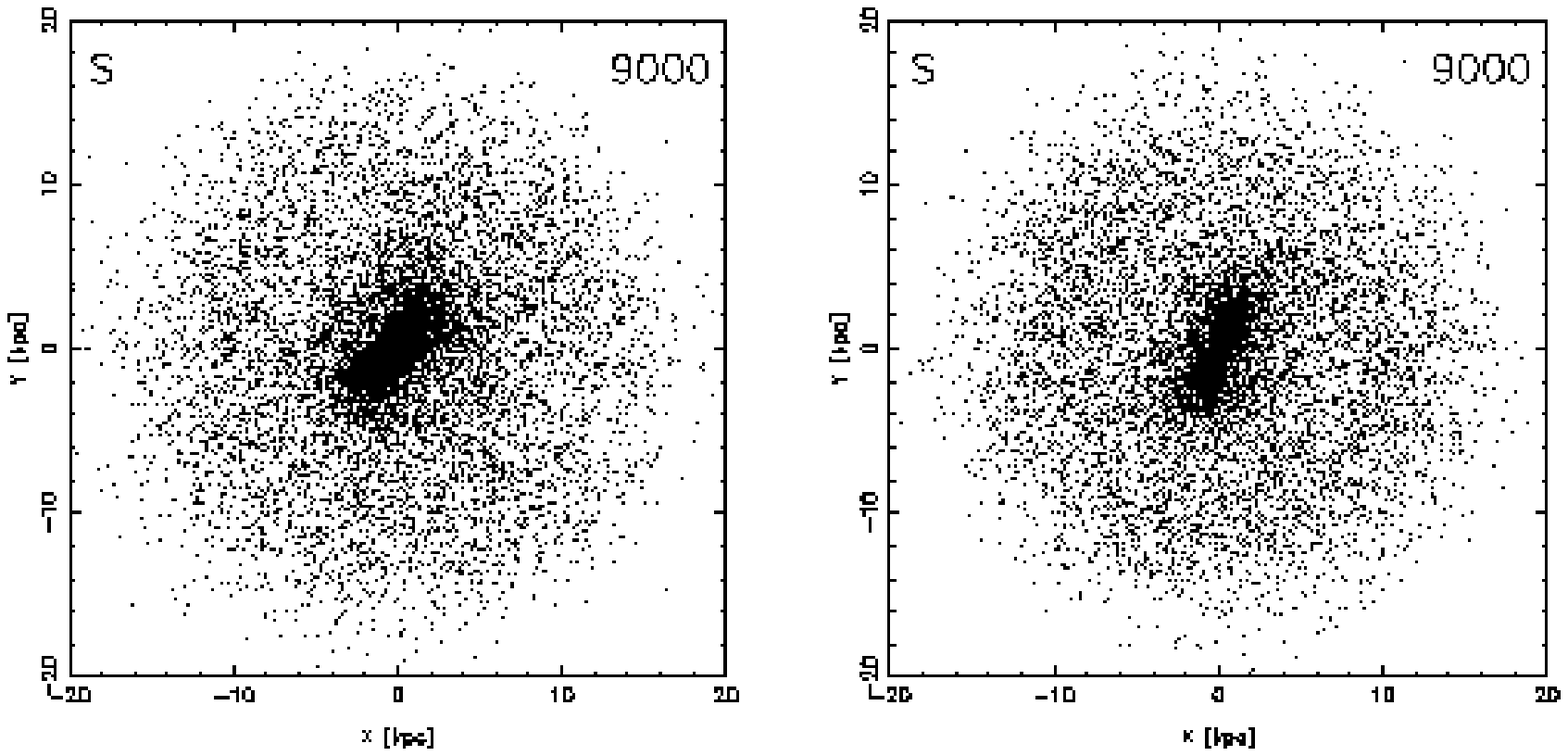}
\vskip5mm
\includegraphics[width=9cm]{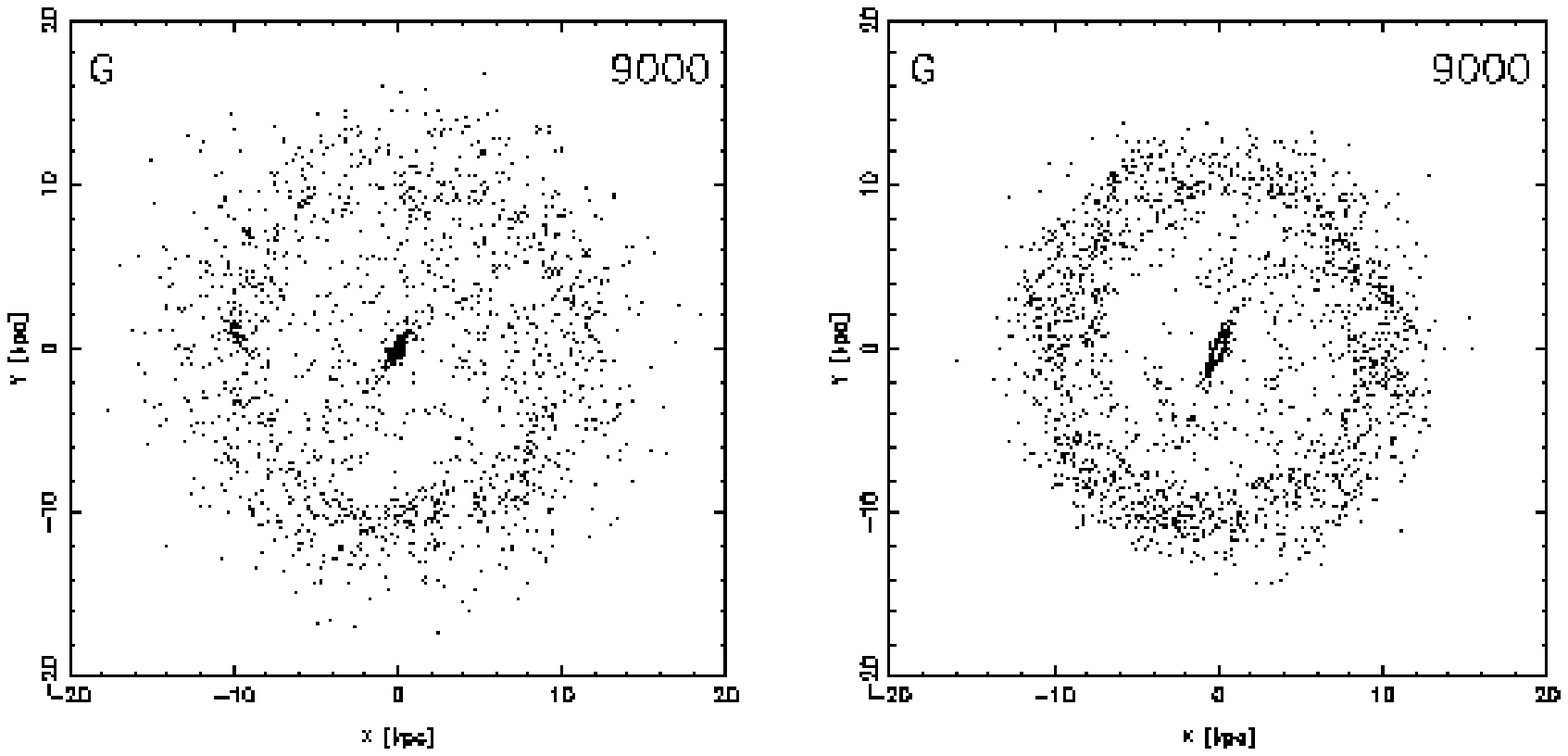}
\vskip5mm
\includegraphics[width=9cm]{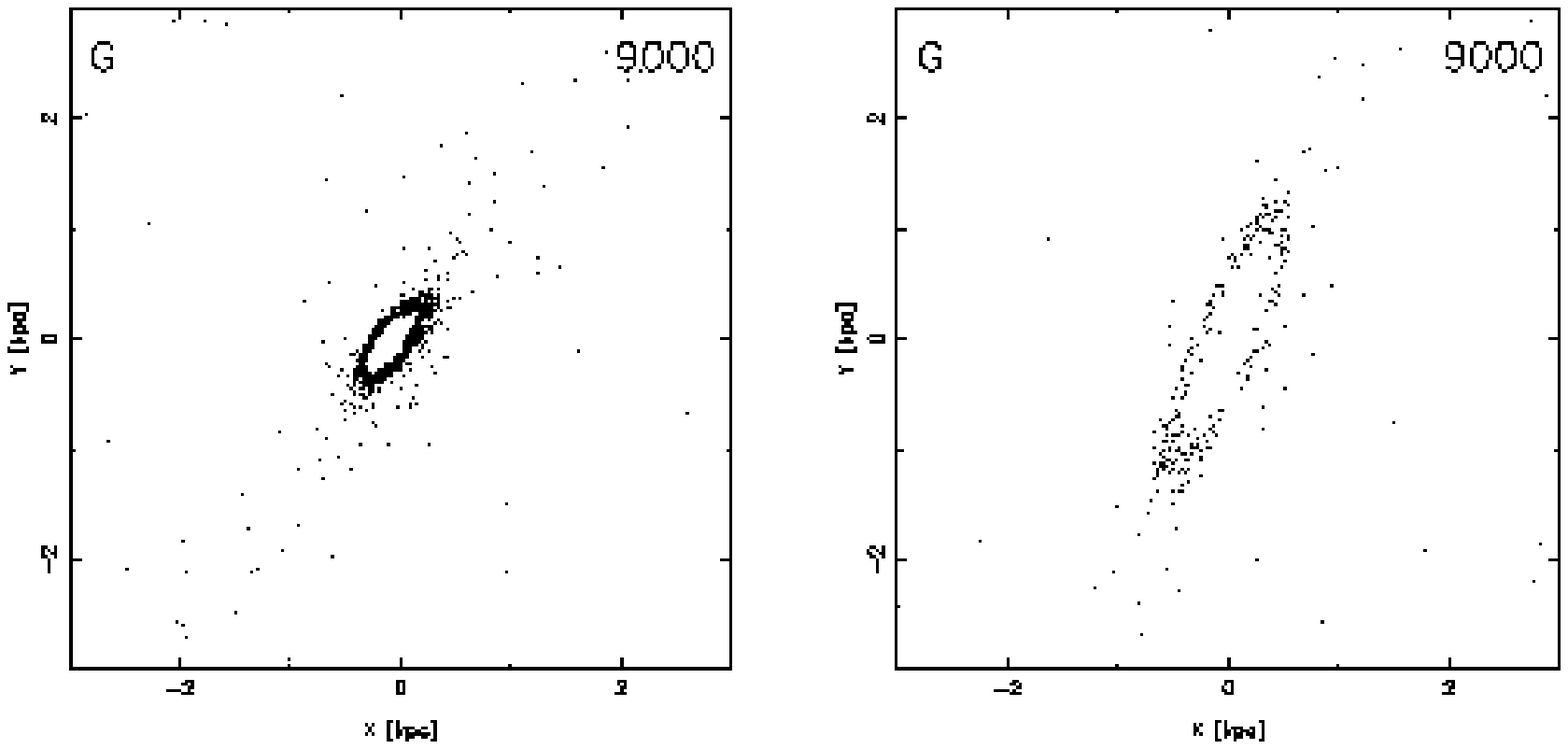}

\vskip3mm
\caption[] {\small Models B (left pannels) and C2 (right pannels)
at $t=9$ Gyr.
{\bf Top}: standard stars; 
{\bf middle}: standard clouds; 
{\bf bottom}: standard clouds -- zoom to the central region.}
\end{figure}

\begin{figure} [htpb]
{\includegraphics[height=4.30cm, width=4.30cm, angle=0]{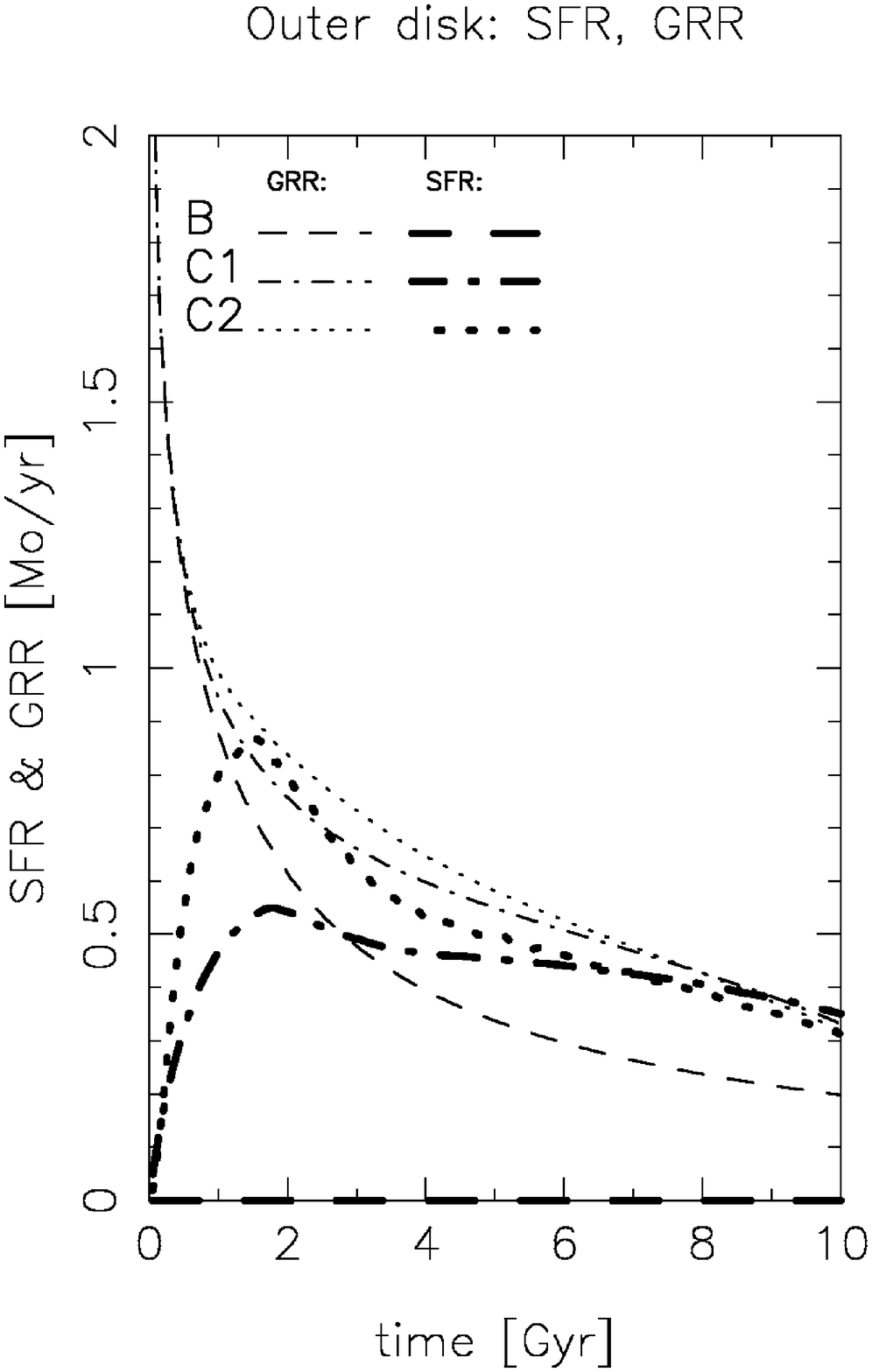}
 \includegraphics[height=4.30cm, width=4.30cm, angle=0]{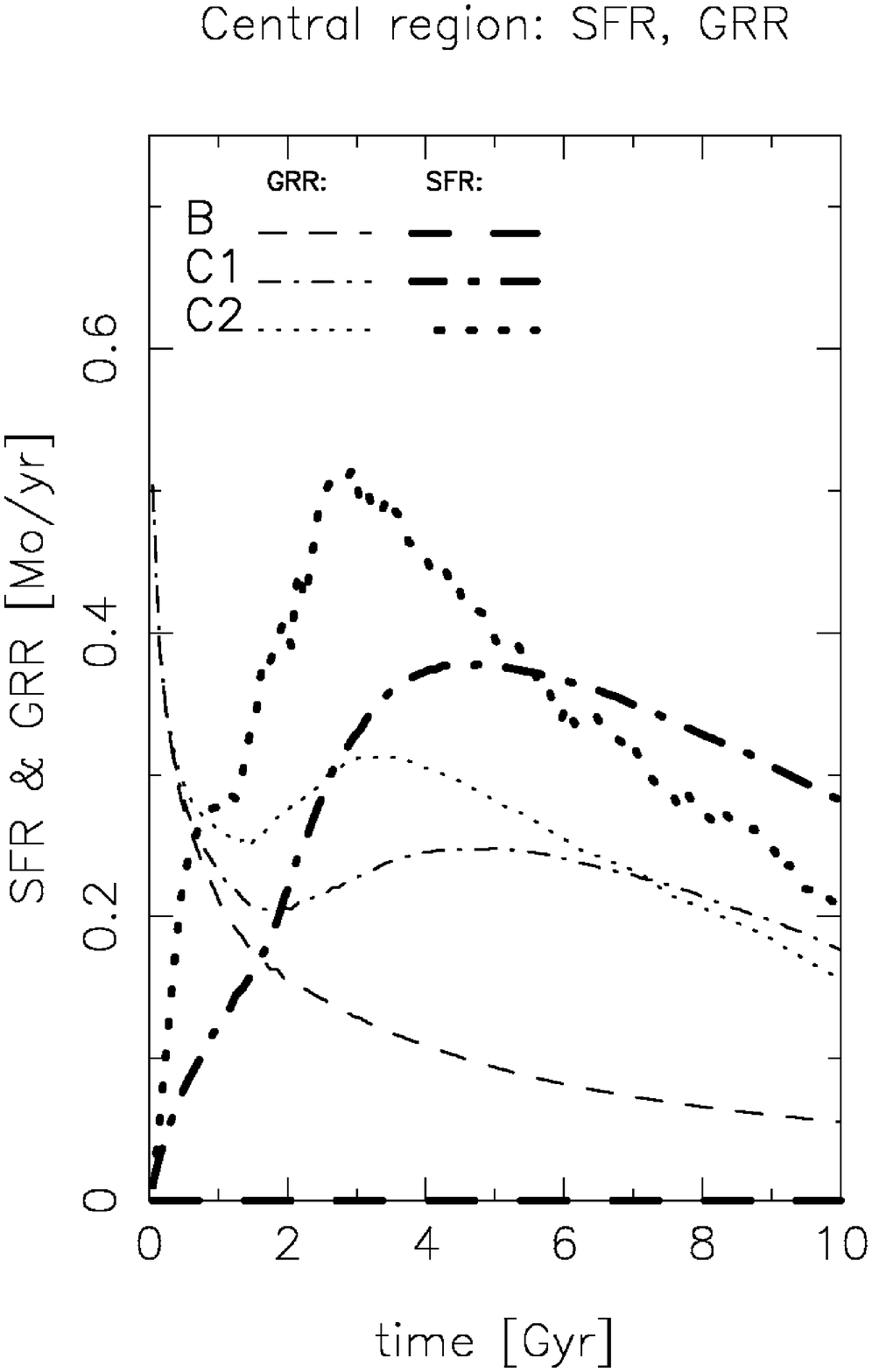}}
\vskip2mm
{\includegraphics[height=4.30cm, width=4.30cm, angle=0]{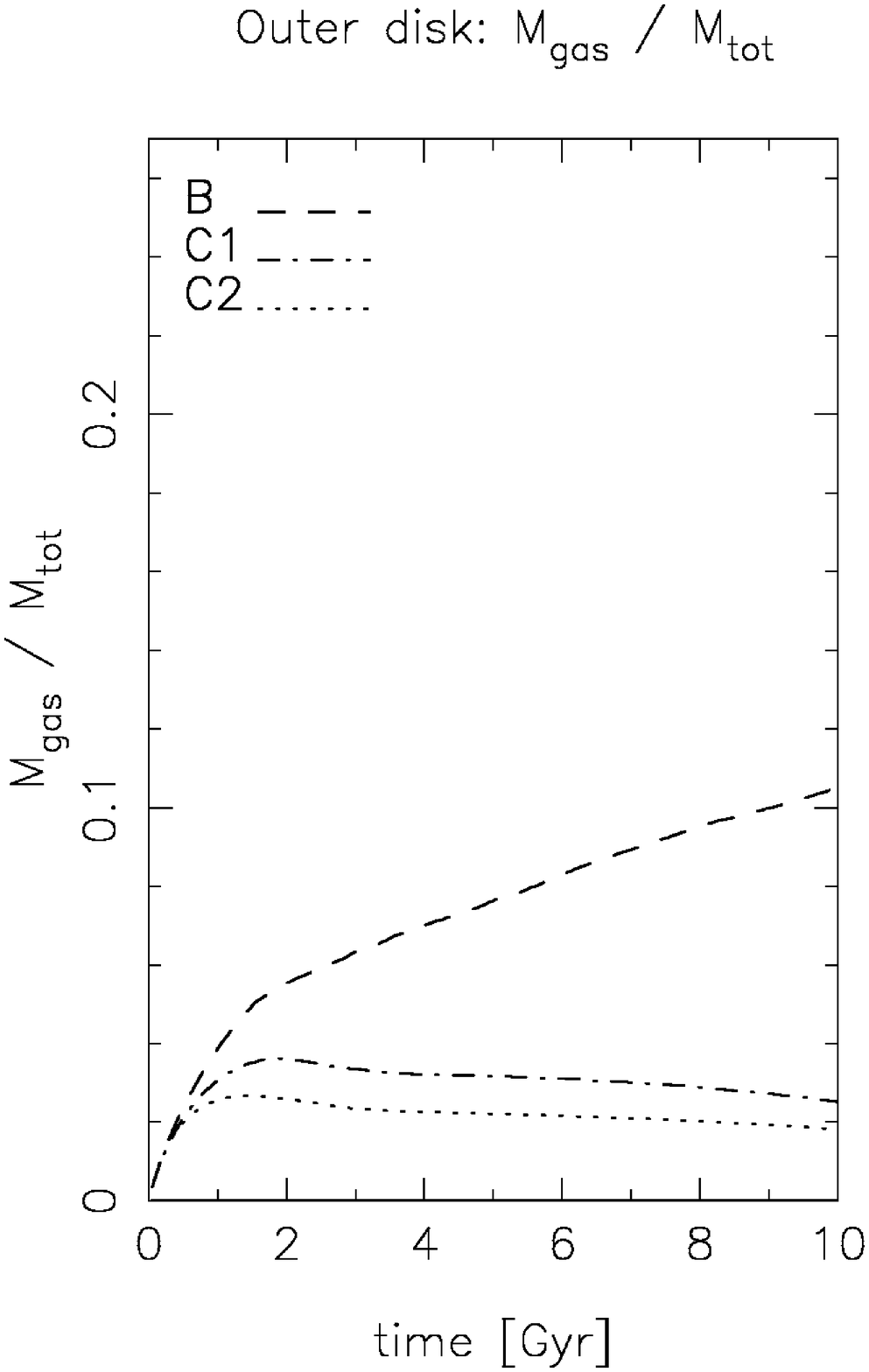}
 \includegraphics[height=4.30cm, width=4.30cm, angle=0]{figure8d.eps}}
\vskip2mm
{\includegraphics[height=4.30cm, width=4.30cm, angle=0]{figure8e.eps}
 \includegraphics[height=4.30cm, width=4.30cm, angle=0]{figure8f.eps}}
\caption[]{\small Models A, B, C1, C2 -- {\bf left}: outer disk ($R>2$ kpc),
{\bf right}: central region ($R<2$ kpc); {\bf top}: $GRR$ (thin lines)
and $SFR$ (thick lines), {\bf middle}: $M_{\rm gas}/M_{\rm tot}$,
{\bf bottom}: $M_{\rm tot}$.}

\vskip5mm
{\includegraphics[height=4.30cm, width=4.30cm, angle=0]{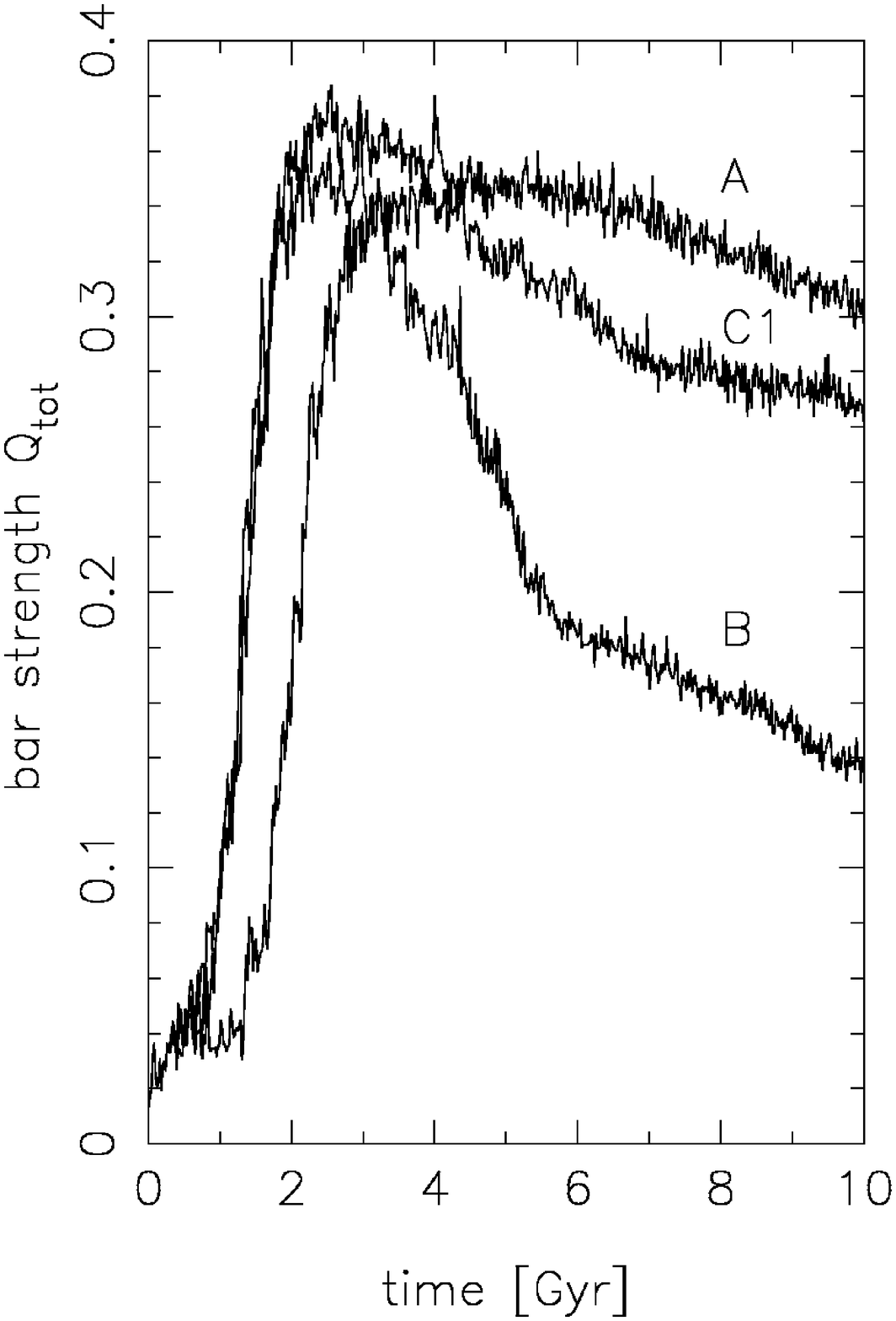}
 \includegraphics[height=4.30cm, width=4.30cm, angle=0]{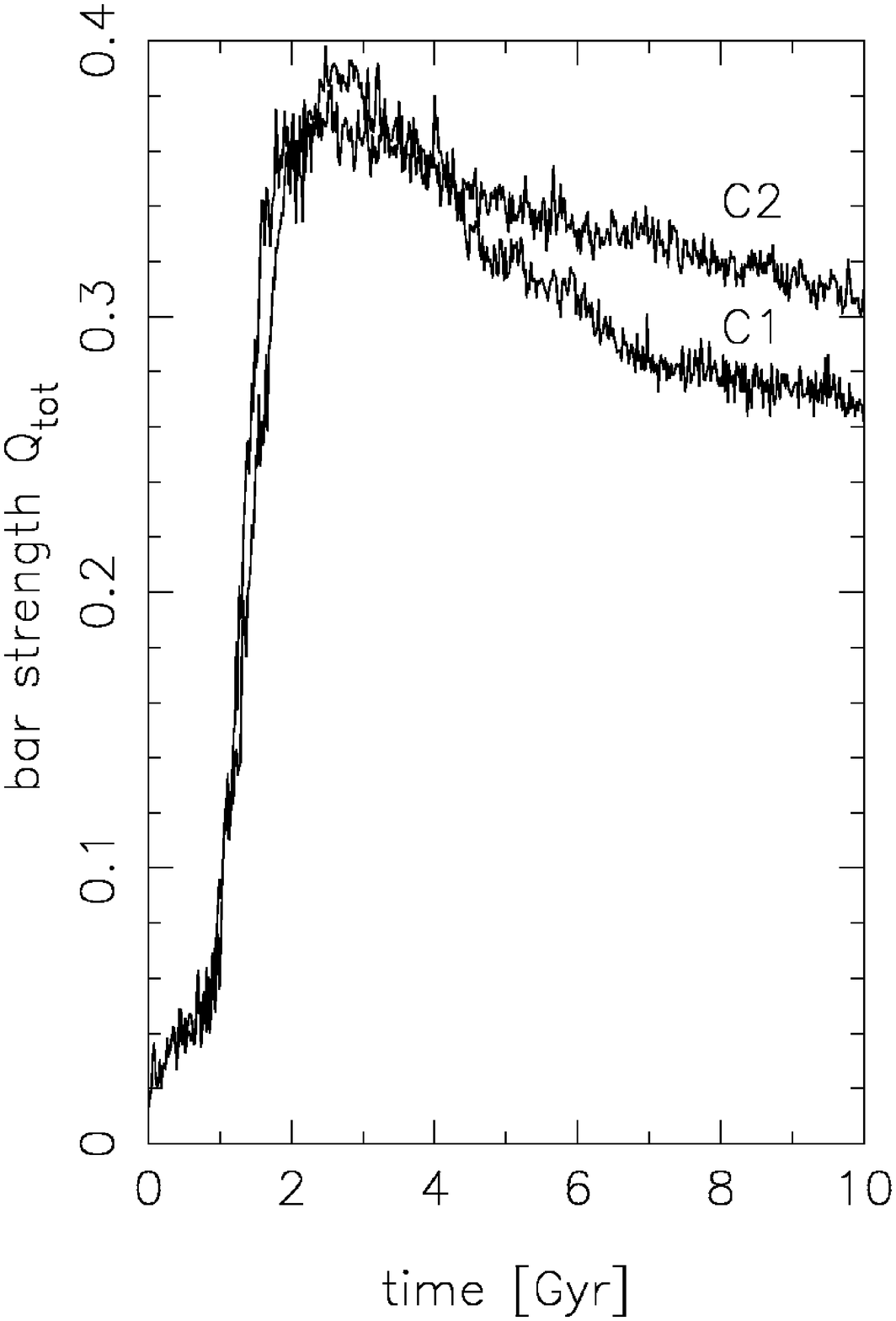}}

\caption[] {\small Bar strength $Q_{\rm tot}$ --
{\bf left}: models A, B and C1;
{\bf right}: models C1 and C2.}
\vskip-5mm
\end{figure}

For models B and C2 (Fig. 7, left and right pannels, respectively), 
we show only an advanced evolutionary stage ($t=9$ Gyr) that can
be compared with the last column snapshots of Figs. 5 and 6.
In model B, one can note that the bar is shorter and rounder (than
in all the other models), the
outer ring is less regular (than in C1 and C2 models) while the nuclear 
ring has collapsed to a smaller size. All these differences are 
related to a huge gas release and associated bar weakening
(for a review, see Combes 2001) that are 
discussed below. The disk structure and ring shapes in model C2 
are rather similar to those of model C1, 
except that the standard (i.e. relatively massive) clouds 
in the nuclear ring are considerably depleted by
a more intense star formation (resulting from a 
higher star formation efficiency at high gas densities).     

As indicated under item 10 of Sect. 4.1., we have chosen
the assumption of constant $SFR$
over 5 Gyr preceding the start of the simulation. Given the initial
disk stellar mass of $4\cdot 10^{10}\, \Mo$, Table 3 implies that
the gas mass released by the pre-existing stars over the simulation
is about $6 \cdot 10^9\, \Mo$. While our choice of the star formation
history is arbitrary and favouring a rather huge mass-loss, Table 3
indicates that other choices would lead to a gas restitution of the same
order.

We show in Fig. 8, separately for the central disk region
(here defined by $R<2$ kpc) and for the outer disk ($R>2$ kpc),
temporal evolutions of $SFR$, $GRR$,
total (i.e. stellar + gaseous mass) and gas mass fraction.
The gas restitution by pre-existing stars in models B, C1 and C2 has
important effects. First, it opens the way
to star formation. Second, the presence of gas affects
the radial transport of matter.

As can be seen in Fig. 8, $SFR$ is of the order of $1\, \Mo\, yr^{-1}$
in the disk as a whole and, in the outer disk,
it is pretty constant over the whole simulation, except early stages.
The relative constancy of $SFR$ is a result of the time-dependent recycling
as was demonstrated by Jungwiert (1998) on a larger set of models, starting
from either a two-component disk with
$M_{\rm gas} / M_{\rm stars} = 0.1$ or from a purely
gaseous disk. As for $GRR$, we note that, except early stages, contributions
of the pre-existing stellar population and of stars formed during
the simulation are comparable (the latter contribution corresponds to
the difference between $GRR$ curves for models C1 (or C2) and for the non-SF
model B).

The transport of matter towards the center is quantified in Table 5 that
shows the mean transfer rate of total (stellar + gaseous)
mass into the central region ($R<2$ kpc)
and the relative strength of the bar, $Q_{\rm tot}$
(defined as the ratio of the maximum bar tangential force over azimuthally
averaged radial force of the total force field; 
Combes \& Sanders 1981; Buta \& Block 2001) at 10 Gyr.

%
\begin{table}
\caption{Mean mass transfer rate (into $R<2$ kpc) and bar strength}
\begin{center}\small
\begin{tabular}{ccc}
      & $<{{\rm d}M_{\rm tot} \over {{\rm d}t}}>_{\rm inflow} 
[\Mo\, yr^{-1}]$ & 
        $Q_{\rm tot}$ (10 Gyr)  \\
Model &  &               \\
A     &  0.17           &   0.30        \\
B     &  0.27           &   0.14        \\
C1    &  0.33           &   0.27        \\
C2    &  0.34           &   0.30       
\end{tabular} \end{center} \end{table}

%
\begin{table}
\caption{Gas mass fractions at 10 Gyr}
\begin{center}\small
\begin{tabular}{ccc}
      &  $M_{\rm gas} (R<2\, {\rm kpc}) \over 
         {M_{\rm tot} (R<2\, {\rm kpc})}$    
      &  $M_{\rm gas} (R<2\, {\rm kpc}) \over 
         {M_{\rm gas} (R<20\, {\rm kpc})}$    \\
Model &  &               \\
B     &  0.260          &   0.46        \\
C1    &  0.055          &   0.45        \\
C2    &  0.006          &   0.11    
\end{tabular} \end{center} \end{table}

The radial tranport is the least efficient in model A, as expected, since
there is no gas. Model B is the most gas rich since mass-loss
operates but star formation is
inhibited, however it is not the one with the largest inflow. The most
efficient are models C1 and C2. This behaviour can be understood with
the help of Fig. 9 that shows evolutions of the bar strength $Q_{\rm tot}$.
In model A, the bar
instability appears after about 1.5 Gyr, the bar strength steeply grows
until about 3 Gyr, then saturates at the value of $\sim 0.33$ and after
5 Gyr enters a phase of only slow secular weakening
(at 10 Gyr, $Q_{\rm tot}=0.30$).
The bar evolution in models B and C differs from model A in two respects.
First, the bar appears earlier (instability starts before $t=1$ Gyr and
the saturation is reached at $\sim 2$ Gyr) since
the gas production helps to destabilize the disk. Second, after saturation,
there is more important weakening of the bar, especially pronounced
in the gas rich model B (bar strength decreases from 0.35 to 0.14).
This is connected to the known fact that a too
important accumulation of matter near the center
disturbs the bar orbital structure 
(Pfenniger \& Norman 1990; Hasan \& Norman 1990; Friedli \& Benz 1993).
The high gas fraction in the disk
of model B thus first leads to a rapidly growing central concentration of gas
(in the nuclear ring) which in turn disturbs the bar thus slowing down
the future mass transfer (see Heller \& Shlosman 1996).

The recycling models C1 and C2 thus provide, from the point of view
of the efficiency in transferring matter inwards, the best combination
of gas mass fraction and bar force, both quantities being
interdependent.

Table 6 gives two gas mass fractions at the end of simulations: 
1) ratio of gas-to-total mass inside the central region (here,
``total'' means gaseous + stellar but without the bulge contribution);
2) ratio of gas masses inside the central region and in the disk as a whole.
In model B, about one quarter of the central region dynamical mass is in the
form of gas. Note also that in models B and C1 roughly one half of the 
gas mass is located in the central region, i.e., more or less in the
nuclear ring.

\subsection{Computational efficiency and statistics of particle species}

The presented algorithm for the continuous stellar mass-loss
has a low computational cost. In the standard model C1, with
50 000 standard stars and active grid of size
256 x 256, the mass-loss and star formation related subroutines
used about 12\% and 3\%, respectively, of the total CPU, 
its main part being
spent on calculating the gravitational potential and advancing the
particles. The simulations were carried out on a relatively
slow one-processor PC Pentium II 366 MHz with one time step,
$\Delta t$, taking on the average 0.9 sec.

\begin{figure} [htpb]
{\includegraphics[width=4.30cm, angle=0]{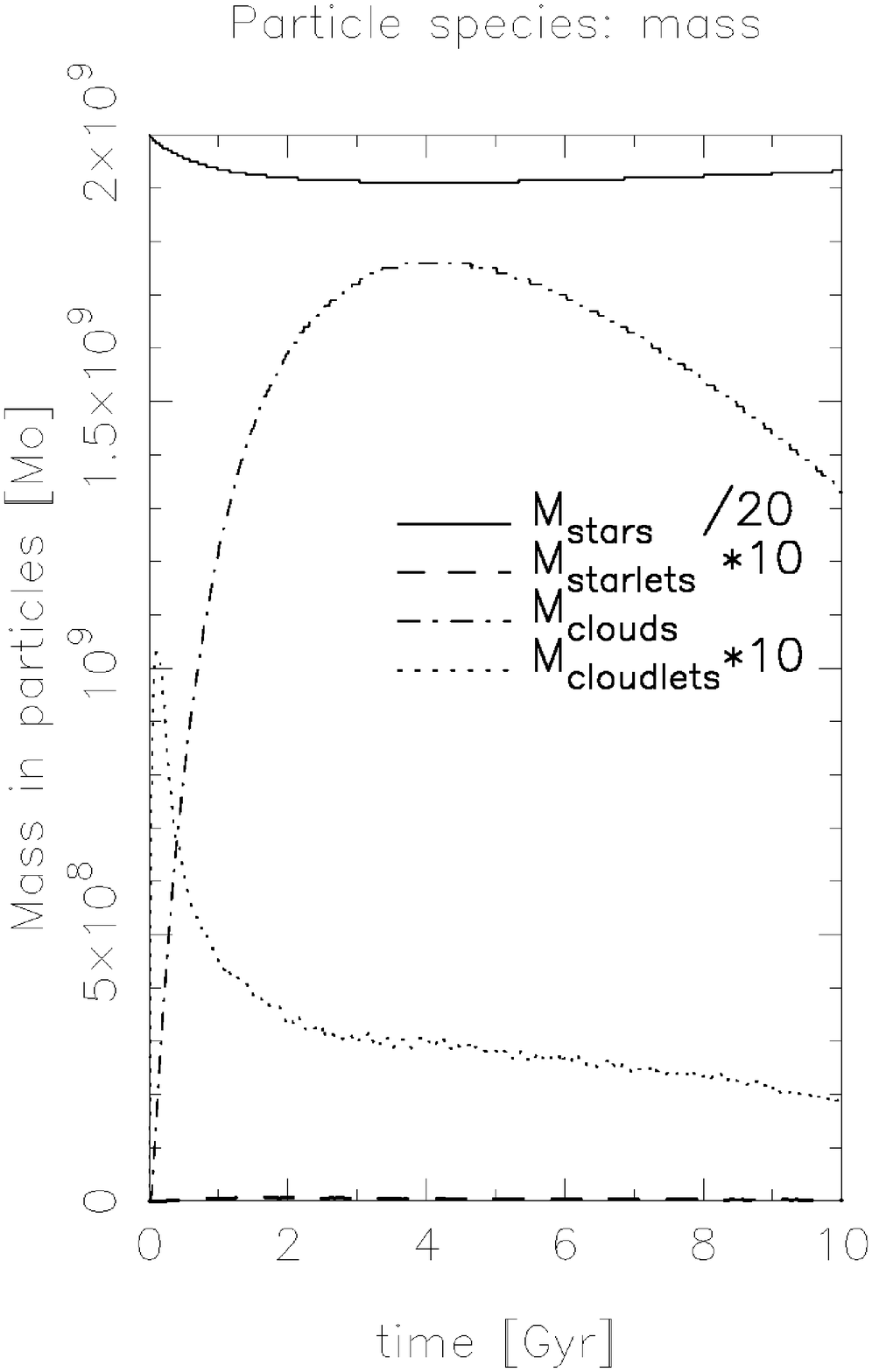}
 \includegraphics[width=4.30cm, angle=0]{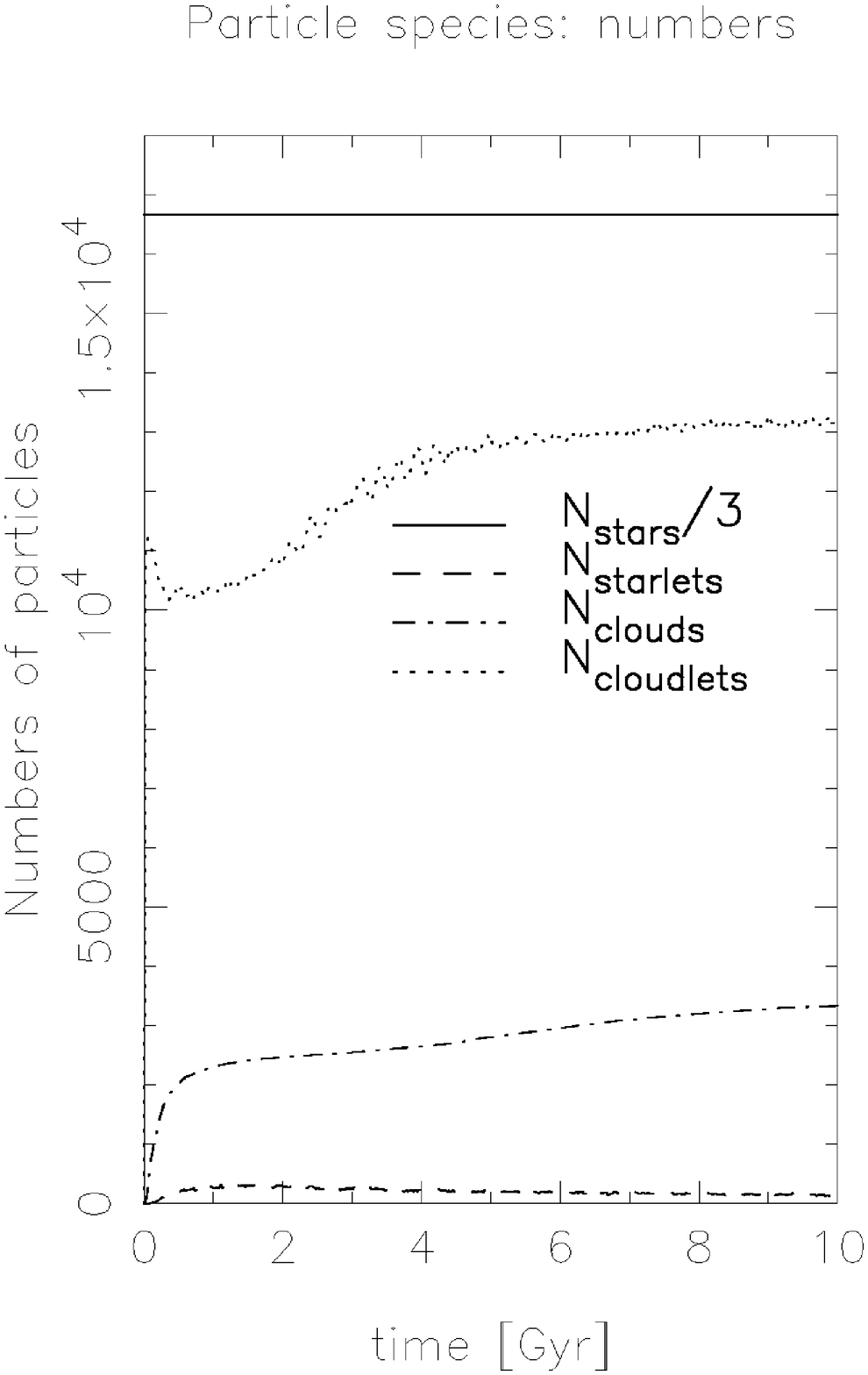}}

\caption[] {\small Masses (left) and numbers (right) of different 
particle species (for convenience, the mass and number of 
standard stars are scaled down by factors of 20 and 3, respectively,
while masses of starlets and cloudlets are scaled up by a factor
of 10.} 
\vskip5mm
\includegraphics[width=9cm, angle=0]{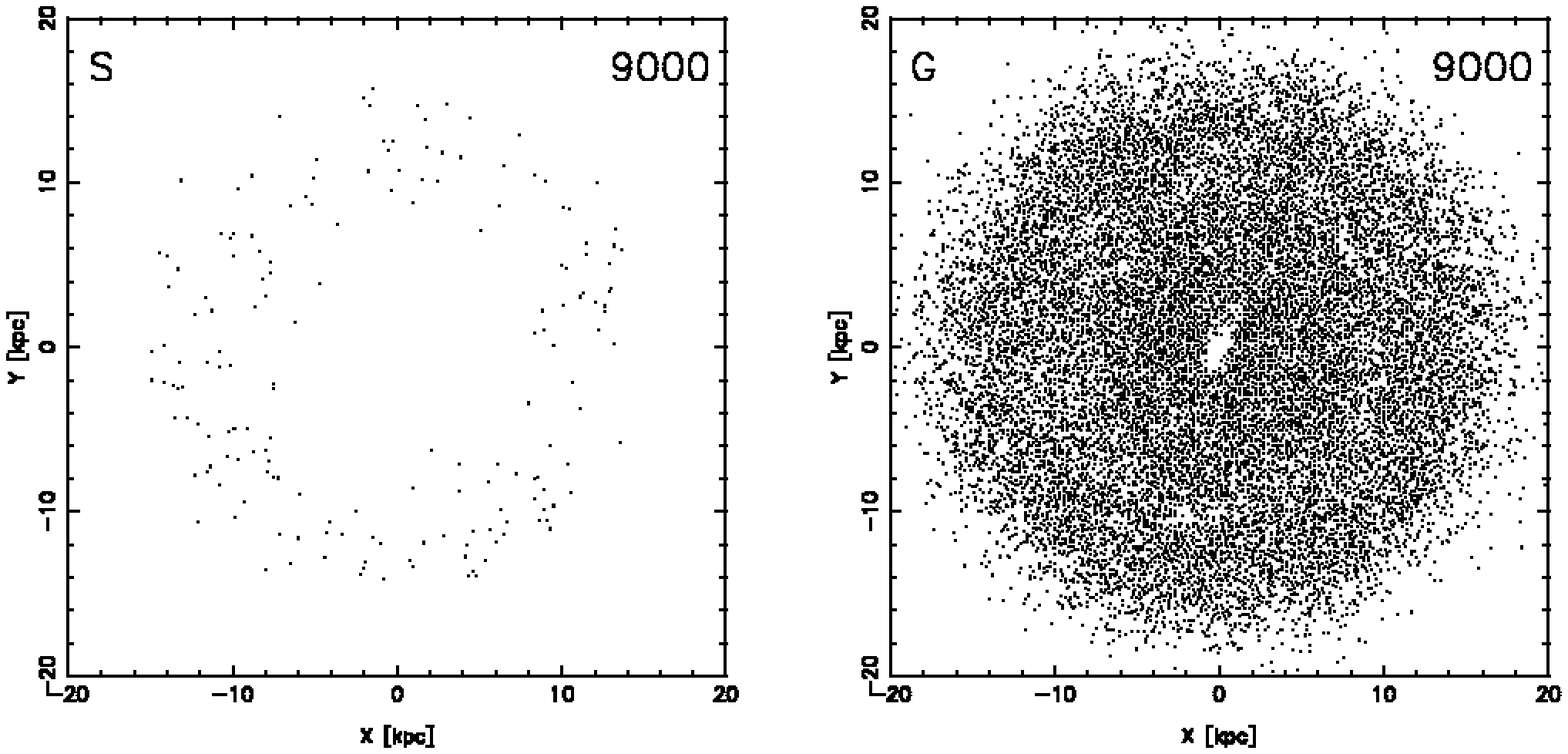}

\caption[] {\small Distribution of starlets (left) and cloudlets (right)
in model C1.}
\vskip-5mm
\end{figure}
 
Fig. 10 gives the statistics (masses and numbers) 
of the four particle species for the model C1. Fig. 11 shows
spatial distributions of starlets and cloudlets.
  
The number of standard stars was constant. In principle, it cannot decrease
since these particles have no way to disappear (except escapes
that however did not occur) but it could rise due to starlets-star
conversion after a starlet reaches the critical mass $M_{\rm sS}$ (Sects.
3.2. and 4.1., item 13); such conversions however did not happen. 
The mass in standard stars slowly varies due to their
mass-loss and absorption of starlets.

The number of standard clouds quickly (on a time-scale of $\sim 500$ Myr)
grows (due to the initially copious stellar mass-loss) 
from $0$ to $\sim 2000$ and then continues climbing only mildly 
(to 3300 at $t=10$ Gyr). On the other hand, the overall 
mass in these particles reaches a maximum ($\sim 1.8 \cdot 10^9\, 
\Mo$) at about 4 Gyr and then declines 
(to $\sim 1.3\cdot 10^9\, \Mo$ at $t=10$ Gyr) 
due to intense star formation and
decreasing mass-loss (cf. Fig. 8).  

The number of starlets and the total mass in them are very low 
throughout the simulation ($\sim 200$ and $10^6 \Mo$) and
negligible compared to standard stars. 
The starlets are thus almost virtual particles: 
despite a big number of them (of the order of 1000; 
one new starlet per box which hosts at least one standard cloud) 
is formed every time step $\Delta t_{\rm ML}$ 
(Sect. 4.1., item 12), the overwhelming majority is immediately
incorporated to nearby standard stars. The only surviving
starlets are located in outer parts of the galactic disk
(Fig. 11, left) where the number of standard stars is low
for the star-starlet mergings to work with the 100\% 
efficiency.

The situation is somewhat different for cloudlets since their
number is relatively high, between 10 000 and 13 000. However, 
their total mass is (except very early evolutionary stages) much
lower (it is scaled up by a factor of 10 in Fig. 10 !) 
than the mass in standard clouds.
Initially, the cloudlets happen to gather $\sim 10^8 \Mo$ 
due to the mass-loss from initial standard stars, 
however they efficiently lose it in favour of the standard
clouds (initially only by cloudlet-cloud conversions, 
later on also by cloud-cloudlet mergings). Cloudlets survive
only in regions of low total gas density and most of the time
harbour only a few $10^7 \Mo$.   
Looking at Fig. 11 (right panel),
one sees that the distribution of cloudlets is largely complementary to that 
of the standard clouds: cloudlets are depleted in high-gas-density regions, 
i.e. in the nuclear and outer gaseous rings. Note also that the cloudlet
distribution extends beyond the outer edge of the standard cloud 
distribution (cf. Fig. 6).

Overall, our star-formation and mass-loss schemes with four particle
species keep the total number of all particles ($\sim 65\, 000$) 
close to the initial value ($50\, 000$) and the various types of the particle
species interactions do not put a heavy load on the CPU time.

\section {Concluding remarks}

We have presented a gas dynamical N-body model including star formation
and time-dependent stellar mass-loss. The implementation of the latter
feature is, both physically and technically, the most innovative element
of our simulations and, as such, is the major result of this work.
The related description should allow other people studying the galactic
evolution by means of N-body simulations to transpose our mass-loss
scheme into their codes.

Our code is obviously not free of shortcomings,
for instance the ignorance of gas
cooling and heating, inherent to sticky-particles representation of gas
(we however recall -- see Sect. 3.2. -- that our subdivision of 
gas particles into standard clouds and cloudlets takes into 
account, on a very rudimentary level, at least some aspects of
the gas cooling and of the gas transition from hot/warm to cold). 
Questions about the mergings of stellar particles were pointed out 
in Sect. 3.2.
Other potentially important simplification
is the neglection of the growth of metallicity on the stellar
mass-loss. 
This will be remedied in the forthcoming research that
will include the chemical enrichment.

We have carried out a simple 2D simulation of a disk galaxy starting
with no gas. Its purpose was 
essentially illustrative. It should demonstrate the functionality
and potential applicability of the developed code. Our intention was also
to show, how the dynamics of classical purely stellar simulations
could be altered if stellar mass-loss were taken into account.

It goes without saying that the choice of special initial conditions
-- on one hand, zero gas mass in the beginning of the simulation and,
on the other hand, presumed constant star formation rate prior to
the simulation -- is not coherent.
Nevertheless, such conditions can be approached  
in some astrophysical situations: during strong interactions between
galaxies, gas is stripped from the outer parts of galactic disks 
(e.g. Moore et al. 1996) while
that falling towards the center can be largely consumed by a 
nuclear starburst (e.g. Mihos \& Hernquist 1996). 
A quietly star forming galaxy can thus be turned into
a state without much gas. It can however be replenished by the stellar 
mass-loss as is the case of our simple model. 
Moreover, our results as for the
rate of mass inflow, enhanced by the time-dependent recycling
(relative to pure N-body calculations),
seem not to qualitatively depend on our special initial conditions 
since they point in the same direction as the results we obtained
for two-component disk models (Jungwiert 1998).

This work will be followed by a second paper dealing with
the application of the presented code to disk galaxies with the
emphasis on comparing the instantaneous recycling approximation with
our time-dependent recycling, especially as to effects on star formation
rate and radial transport of matter.

\begin{acknowledgements}

This research was carried out in the framework of the Czech-French project
No. 9902 that is co-financed by the Academy of Sciences of the
Czech Republic (AV \v CR) and by the Centre National de la Recherche
Scientifique (CNRS) and in the framework of the Key Project K1048102
of AV \v CR. It was also partly supported by the grant No. A3003705 of
the Grant Agency of AV \v CR.
We thank the anonymous referee for careful
reading of the manuscript and for his suggestions.

\end{acknowledgements}

\vfill\eject

\end{document}